\documentclass{aa}

\usepackage{graphicx}
\usepackage{txfonts}
\usepackage{xcolor}
\usepackage{hyperref}
\definecolor{linkcolor}{rgb}{0.390625,0.5607843137,0.99609375}
\hypersetup{
  linkcolor  = linkcolor,
  citecolor  = linkcolor,
  urlcolor   = linkcolor,
  colorlinks = true
}
\hypersetup{
  linkcolor  = linkcolor,
  citecolor  = linkcolor,
  urlcolor   = linkcolor,
  colorlinks = true
}
\usepackage{graphicx}
\usepackage{amsmath}
\usepackage{siunitx}
\usepackage{braket}

\usepackage{bm}
\usepackage{mathtools}
\usepackage{bbold}
\usepackage{fontawesome}
\usepackage{graphicx}
\newcommand*\Laplace{\mathop{}\!\mathbin\bigtriangleup}
\newcommand*\dd{\!\mathop{}\mathrm{d}}
\DeclareSIUnit \h {\ensuremath{\mathit{h}}}
\DeclareSIUnit \parsec {pc}
\DeclareSIUnit\year{yr}

\DeclareMathOperator*{\argmax}{arg\,max}

\makeatletter
\renewcommand*\aa@pageof{, page \thepage{} of \pageref*{LastPage}}
\DeclareRobustCommand{\HI}{%
  \mbox{H\check@mathfonts\fontsize\sf@size\z@\selectfont I}%
}
\makeatother

\begin{document}

   \title{Interference in Fuzzy Dark Matter Filaments}
   \subtitle{Idealised Models and Statistics}

   \author{%
       T. Zimmermann\inst{1},
       D. J.E. Marsh\inst{2},
       K. K. Rogers\inst{3,}\inst{4},
       H. A. Winther\inst{1},
       \and
       S. Shen\inst{1}
    }

   \institute{%
        Institute of Theoretical Astrophysics, 
        University of Oslo, PO Box 1029, Blindern 0315, Oslo, Norway\\\email{timzi@uio.no}
    \and
        Theoretical Particle Physics and Cosmology, King’s College London, Strand, London WC2R 2LS, UK
    \and
        Department of Physics, Imperial College London, Blackett Laboratory, Prince Consort Road, London, SW7 2AZ, UK
    \and
        Dunlap Institute for Astronomy and Astrophysics, 
        University of Toronto, 50 St. George St., Toronto, ON, M5S 3H4, Canada
   }
   \date{Received XXXXXX XX, 2025; accepted XXXX XX, XXXX}


  \abstract{%
        Fuzzy (wave) dark matter (FDM), the dynamical model underlying an ultralight
        bosonic dark matter species, produces a rich set of
        non-gravitational signatures that distinguishes it markedly from
        the phenomenologically related warm (particle) dark matter (WDM) scenario.
        The emergence of extended interference fringes hosted by cosmic filaments is one such phenomenon reported by cosmological simulations, and a detailed
        understanding of such may strengthen existing limits on the boson mass but
        also break the degeneracy with WDM, and provide a unique fingerprint of interference in 
        cosmology.
        In this paper, we provide initial steps towards this goal. In particular,
        we show in a bottom-up approach, how the presence of
        interference in an idealised filament population can lead to a
        non-suppressive feature in the matter power spectrum --- an observation
        supported by fully-cosmological FDM simulations.
        To this end, we build on a theoretically motivated and numerically
        observed steady-state approximation for filaments and express
        the equilibrium dynamics of such in an expansion of FDM
        eigenstates. We optimize the size of the expansion by incorporating
        classical phase-space information. Ellipsoidal collapse considerations
        are used to construct a fuzzy filament mass function which, together with the
        reconstructed FDM wave function, allow us to efficiently compute the
        one-filament power spectrum.
        We showcase our non-perturbative interference model for a selection of
        boson masses and confirm our approach is able to
        produce the matter power boost observed in fully-cosmological FDM
        simulations. More precisely, we find an excess in correlation between
        the spatial scale associated with the FDM ground state and the quantum
        pressure scale. We speculate about applications of this effect in data analysis.
  }


   \maketitle
\section{Introduction}
If the observed dark matter (DM) in the Universe~\citep{Planck2018} is bosonic and has a sufficiently 
low mass, $m\lesssim 1\text{ eV}$, then it can be modelled as a classical wave on astrophysical scales. 
For definiteness, we consider a scalar (or pseudoscalar) field. The gravitational physics of a 
scalar field is distinct from a gas of cold, pressureless particles i.e. cold DM (CDM). 
Scalar fields obey the Klein-Gordon equation, while CDM obeys the collisionless Boltzmann equation. 
If the particle mass is sufficiently low, $m\lesssim 10^{-18}\text{ eV}$, then the distinction between 
wavelike DM and CDM can show up on cosmological scales. Since the effect of DM waves is typically to 
smooth out structures relative to CDM, such a model is referred to as 
``fuzzy'' DM \citep[FDM, see e.g.][] {Hu2000,Marsh2014,Marsh2016,Hui2017}. 

If all the DM is to be one fuzzy particle, then its mass is bounded from below to be 
$m\gtrsim 10^{-21} - 10^{-19}\text{ eV}$ with the weaker limit coming from
Milky Way satellite galaxy population statistics and their internal structure ~\citep{Nadler2019,Zimmermann2024}, 
and the stronger limit 
coming from dynamics of star clusters in ultrafaint dwarf
galaxies~\citep{Marsh2019,Dalal2022}. An array of limits populates the parameter space in between \citep{Irsic2017,Kobayashi2017,Armengaud2017,Nori2018b,Nadler2019,Rogers2021,Banik:2019smi,Powell2023}.
FDM of lower mass $10^{-32}\text{ eV}\lesssim m\lesssim 10^{-24}\text{ eV}$ is restricted to be a 
sub-dominant component of the DM making up at most 20\% of the total mass 
density~\citep{Hlozek2015,Hlozek2018,Kobayashi2017,Lague2021,Dentler:2021zij,Rogers2023,Winch2024}. 
The prevalence of such ultralight particles in high energy physics 
models~\citep{Arvanitaki2010,Mehta2021,Cicoli2021, Sheridan2024}, together with their potential of alleviating tensions between cosmological and astrophysical probes \citep{Rogers2023,Rogers2025},
encourages the search for 
cosmological evidence of FDM across the entire allowed parameter space. 

The wavelike nature of FDM leads to a number of physical effects distinct from 
CDM~\citep{Marsh2021}, including the existence of a Jeans scale~\citep{Khlopov1985}, 
the formation of solitons~\citep{Schive2014,Levkov2018}, and turbulence leading to 
relaxation~\citep{Hui2017}, and all of these effects have been exploited in efforts to 
constrain and search for FDM. There is no more canonical signature, however, that a particle is in fact 
a wave than the presence of interference fringes. In fully cosmological simulations of FDM interference 
fringes are observed in a striking way inside 
\emph{cosmic filaments}~\citep{Schive2014,Mocz2019,May2022}. As noted by 
\cite{Mocz2019}, the presence of such interference is a smoking gun of FDM that distinguishes 
it markedly from the related warm DM (WDM) model: both WDM and FDM suppress
structure growth on small spatial scales, but only FDM introduces interference patterns
that modulate the density on scales comparable to the (reduced) de Broglie-wavelength
$\lambda_{dB}=\hbar/(mv)$ \citep{Hui2017}. 

Filaments are particularly intriguing structures in the context of constraining
the nature of DM. Particle DM simulations with
and without suppressed initial conditions suggest filaments to realise a volume
filling fraction of $\sim 20\%$ for $z > 3$ while maintaining a fairly constant mass filling
fraction of $\gtrsim 50\%$ \citep{Dome2023b}. Weak gravitational lensing is an
ideal probe for detecting this filamentary DM density \citep{Mead2010} and
observational advances using this detection method are encouraging
\citep{Xia2020,HyeongHan2024}. Lyman-alpha (Ly$\alpha$) emission from neutral hydrogen (\HI{})
tracing DM is another detection mechanism and future
observational facilities, such as the Extremely Large Telescope, may be equipped
to reveal Ly$\alpha$ emitting filamentary structures of the cosmic web
\citep{Liu2024}. Ly$\alpha$ absorption, more precisely the spectrally resolved forest of absorption lines, has proven to be an indispensable probe for
the small-scale behavior of DM \citep[e.g.,][]{Viel:2005qj,Rogers2021,Rogers:2021byl,Irsic:2023equ}. Its flux power spectrum arises from low \HI{} column density regions, tracing less non-linear DM overdensities $<10$ \citep{Lukic2014} common for void-like regions and filament outskirts.
Returning to Fig. \ref{fig:showcase_crosssection}, we find this region to be strongly modulated by interference. In this work, we concretely assess whether the scales on which interference manifests in filaments are above or below the pressure filtering scale of the intergalactic medium that sources the Ly\(\alpha\) forest and thus is relevant for these analyses.

In the context of FDM, we note that in comparison to haloes, filaments are less compact objects, with
shallower gravitational wells and thus smaller characteristic
velocities \citep{Mocz2019}. As wave effects manifest around the de-Broglie scale
with coherence time $t \sim \lambda_\text{dB}/v$, 
interference features at fixed mass are expected to be more pronounced and less
transient compared to highly nonlinear haloes --- an expectation supported by
simulation \citep{Mocz2019, May2022}.

Providing a more principled, theoretical understanding of interference in cosmic
filaments is thus a timely contribution with the ultimate aim to provide a
large scale structure, interference-only DM mass limit.
In this paper we focus on (i) the construction of a simplified model for FDM
interference in filaments and (ii) the question of how to measure interference features statistically.

Fully non-linear simulations of the non-relativistic limit of
Klein-Gordon --- the Schr\"odinger-Poisson equation (SP) --- are challenging
\citep{Schive2014,Mocz2019,May2021, May2022,Lague2023} and approximate treatments remain crucial to
forward model isolated elements of the FDM phenomenology. If one is interested in
large-scale morphological changes of a fuzzy cosmic web and its
constituents, encoding FDM's suppression physics in the initial conditions, but
treating it dynamically as CDM, has proven effective \citep{Dome2023, Dome2023b}.
Combining this ``classical FDM'' approximation with emulation techniques \citep{Rogers2018, Rogers2020}
enabled \cite{Rogers2021} to derive $m>\SI{2e-20}{\electronvolt}\;$ due to a lack
of structure suppression in the Ly$\alpha$ forest.
The emergence of solitonic cores and their relation to its environment, on the other hand, may be faithfully
reproduced by an
approximate hydrodynamic formulation of SP \citep{Mocz2015,Nori2018,Nori2021}.
However, none of these approximate methods are able to produce interference. 
The approximate Gaussian beams method is exceptional in that it can produce
interference~\citep{Schwabe2022}, but it has not been deployed in large studies.
An intriguing alternative to these SP-simulation-based approaches is a physically-informed post-processing 
of non-FDM data. Put simply, in the context of filaments, we seek a self-consistent prescription for 
``painting on interference fringes'' on CDM filaments. This work makes initial steps toward this goal.

Building upon advances in the steady-state modelling of spherically symmetric FDM haloes
\citep{Lin2018, Yavetz2022, Zimmermann2024}, we confine ourselves to
isothermal, steady-state filaments \citep{Stodolkiewicz1963,Ostriker1964,Ramsoey2021}. We translate
the spherically symmetric halo prescription to cylindrical filaments and further
develop the model by incorporating classical phase-space information.

Fully-fledged SP simulations also provide guidance on how interference is expected
to manifest quantitatively. More precisely, \cite{Mocz2019,May2021,May2022}
independently report a
boost in the matter power spectrum at scales $k \ge \mathcal{O}(100)\;h \text{
Mpc}^{-1}$, compared to the smoothed WDM spectrum and ultimately the CDM case at
$z \le 7$. Quite intuitively, it was conjectured that this feature may be related to
interference fringes on scales above the quantum pressure scale. Anticipating
our result, we show in a proof-of-concept analysis that our idealised filament
model is indeed able to produce interference fringes in accordance with these scales and leading to the
power boost fully non-linear simulations suggest.

The paper is organised as follows:
Sec.~\ref{sec:method} motivates, develops, and showcases our idealised toy model
for FDM interference.
Sec.~\ref{sec:statistics} investigates statistical aspects of a fuzzy filament
population and, in particular, its population statistics (i.e. a FDM filament
mass function) as well as two-point density correlation (i.e. the matter power spectrum).
We discuss our findings in Sec.~\ref{sec:discussion}. Conclusions are drawn
in Sec.~\ref{sec:conclusion}.

\section{
    An Idealized Model for Interference in Cosmic Filamants
}
\label{sec:method}
\begin{figure}
    \centering
    \includegraphics[width=\columnwidth]{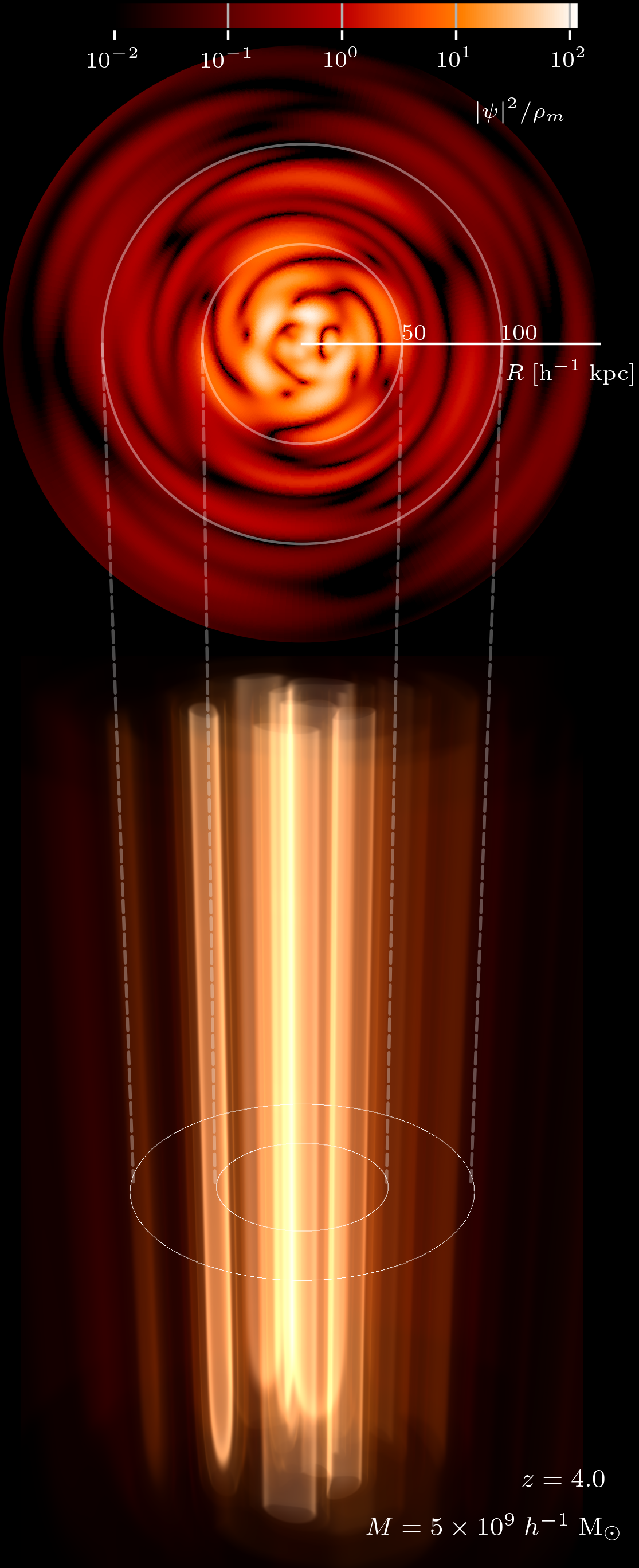}
    \caption{\label{fig:showcase_crosssection}
        \textbf{(Top)} Cross section of a steady-state FDM filament with FDM 
        mass $m=\SI{2e-22}{\electronvolt}$ at redshift $z=4$. 
        The interference is constructed as a post-processing step for an interference-free background density
        which may be provided by an analytical model (this work) or DM
        simulations (future work). The depicted situation assumes isotropy in
        the cross sectional plane. However, our model is designed to apply to
        anisotropic scenarios as well, see Sec.~\ref{sub:filament_cylinder}.
        \textbf{(Bottom)} Volume rendering of a toy filament obtained by
        extruding above cross section along a straight filament spine. Section
        \ref{sec:method} models FDM filaments as infinitely long versions of
        such extrusions. 
        In Sec.~\ref{sec:statistics} we find a physically-sound extrusion length
        for our 2D model via ellipsoidal collapse considerations.
    }
\end{figure}
The point of departure is a discussion of our idealised, semi-analytical model for
interference in FDM filaments --- 
an extension of the regression approach developed in
\cite{Yavetz2022} by a physically-informed prior based on a \emph{steady-state} phase space
information \citep{Widrow1993, Lin2018, Dalal2021}. We refer to Fig.~\ref{fig:showcase_crosssection} for a preview.

After introducing interference reconstruction in a general setting in Sec.~\ref{sub:interference}, we specialise to the case of steady-state filaments 
in Sec.~\ref{sub:filament_cylinder}.
To this end, we develop a self-consistent description of quasi-virial filaments
in both real space, Sec.~\ref{subsub:real_space}, and phase space, Sec.~\ref{subsub:phase_space}. 
Specifics of our modified regression approach are outlined in Sec.~\ref{sub:reconstruction}. 
We close by showcasing our model for a range of relevant FDM masses and
phase space model parameters in Sec.~\ref{sub:interference_in_filaments}.

\subsection{\label{sub:interference}%
    Self-consistent FDM Interference in Steady-State Systems
}
In the non-relativistic limit applicable for the study of cosmic structure formation, 
FDM is governed by the Schr\"odinger-Poisson equation \citep[e.g.][]{Hui2021b}. 
With $m$ denoting the FDM mass, $a(t)$ the scale factor and $G$ as Newton's
constant, the Schr\"odinger-Poisson equation reads:
\begin{subequations}
\begin{align}
    \label{eq:nlse}
    i\hbar\partial_t\psi &=\left[- \frac{\hbar^2}{2m a^2(t)}\Laplace + mV(\bm x, t)
    \right]\psi \;,\\
    \label{eq:fdm_poisson}
    \Laplace V &= \frac{4\pi G}{a(t)}\left(|\psi|^2 - 1\right) \;.
\end{align}
\end{subequations}
$\psi\in\mathbb{C}$ represents the FDM field, commonly dubbed the wave function, which is
self-consistently coupled to its own gravitational potential $V$ via Poisson's
equation. $|\psi|^2$ measures the FDM density in a comoving
volume. For the remainder of this paper comoving coordinate components are denoted with
capital letters while physical components are lower case, e.g. $\bm x = (R, \Phi,
Z)$ for comoving and $\bm r = (r, \phi, z)$ physical cylindrical coordinates.

Equation~\eqref{eq:nlse}, a nonlinear Schr\"odinger equation with explicitly
time-dependent Hamiltonian, is challenging in its 
numerical treatment as all physically relevant scales --- from \si{\mega \parsec}-sized
filaments down to the de-Broglie scale inside haloes ($\lambda_{\text{dB}} =
\hbar/(m\sqrt{\sigma^2}) < \SI{1}{\kilo\parsec}$ with $\sigma^2$ as
characteristic halo velocity dispersion) need to be 
resolved in order to guarantee convergence and stability \citep{Schive2014,Garny2020, May2021}.

Fortunately, if one is only interested in a self-consistent treatment of
interference effects in isolated objects,
it is possible to forego a fully-fledged integration of eqs.~\eqref{eq:nlse}-\eqref{eq:fdm_poisson} under the assumption that the detailed
dynamics encapsulated in $\psi(\bm x, t)$ take place in a smooth gravitational potential
that is effectively static in time \citep{Lin2018, Dalal2021}. 

A canonical example where this holds true is a virialised DM halo.
Violent relaxation and phase mixing \citep{Lynden-Bell1967} drive collisionless 
CDM into a virial equilibrium with universal Navarro-Frenk-White (NFW) density profiles \citep{Navarro1996}. 
The associated relaxation time scale may be approximated with the free fall
time, $t_\mathrm{rel} \approx \left(G\bar\rho\right)^{-1/2}$
\citep{Lynden-Bell1967}, and is characteristic for changes in the potential due
to out-of-equilibrium density configurations.
Similarly, FDM haloes realise NFW-like densities comprised of short-lived
interference granules with coherence time \citep{Hui2021}:
\begin{equation}
    t_c \equiv \frac{\lambda_\text{dB}}{2\sqrt{\sigma^2}} 
    \simeq
    \SI{1}{\mega\year}
    \left(\frac{\SI{1e-22}{\electronvolt}}{m}\right)
    \left(\frac{\SI{250}{\kilo\meter\second^{-1}}}{\sqrt{\sigma^2}}\right)^2 \;.
\end{equation}
This large-scale equivalence to CDM is a consequence of the Schr\"odinger-Vlasov correspondence
\citep{Uhlemann2014, Mocz2018} and it is only at small radii where gravitational Bose-Einstein condensation 
\citep{Levkov2018} causes a deviation due the formation of a solitonic core \citep{Schive2014}. 

For mean halo densities of $200$ times the background matter density, $\bar\rho \approx 200
\rho_m$, we find $t_c \ll t_\mathrm{rel} \ll t_H \approx \left(G \rho_m\right)^{-1/2}$. 
Consequently, one expects the gravitational potential to
appear roughly static on the dynamical time scale of the interference evolution and approximately 
maintained on the time scale of the expanding background. The latter implies
that the scale factor can be treated as fixed.
Note that the halo density and potential enjoy approximate spherical symmetry.

We investigate how similar conditions, i.e. a quasi-static gravitational
potential with a high degree of symmetry may be justified in a limiting case of cosmic filaments in Sec.~\ref{sub:filament_cylinder}.

With a time-independent, interference-free potential in place
eqs.~\eqref{eq:nlse}-\eqref{eq:fdm_poisson} reduce to:
\begin{subequations}
\begin{align}
    \label{eq:se}
    i\hbar\partial_t\psi &=\left[- \frac{\hbar^2}{2m a^2}\Laplace + mV(\bm x)
    \right]\psi \;,\\
    \label{eq:poisson}
    \Laplace V &= \frac{4\pi G}{a}\rho_\text{BG}(\bm x) \;,
\end{align}
\end{subequations}
and eq.~\eqref{eq:poisson} is solved only once to fix the
time-independent Hamiltonian.
Importantly, eq.~\eqref{eq:se} is decoupled from eq.~\eqref{eq:poisson} and
no non-linearity is present. We may interpret this system of PDEs
as the linear order perturbative treatment of
interference with $V$ as zeroth-order potential \citep{Dalal2021}. 

Integrating eq.~\eqref{eq:se} in time reduces to diagonalising its Hamiltonian and
expanding $\psi(\bm x, t)$ in the eigenbasis $\{\psi_j(\bm x)\}_{j=1\dots J}$ with
energy eigenvalues $\{E_j\}_{j=1\dots J}$. The time evolution is thus set by:
\begin{equation}
    \label{eq:evolution}
    \psi(\bm x, t) = \sum_j a_j \psi_j(\bm x) e^{i E_j t / \hbar}\quad\text{with}\quad a_j \in \mathbb{C} \;,
\end{equation}
and the sought-after interference emerges as the cross-terms in the 
\emph{time-dependent} density:
\begin{equation}
    \label{eq:interference}
    |\psi(\bm x, t)|^2  
    = \sum_j |a_j|^2|\psi_j(\bm x)|^2 
    + \sum_{j\neq k} a_j a_k^* \psi_j(\bm x) \psi_k^*(\bm x) e^{i(E_k
    -E_j)t/\hbar} \;.
\end{equation}

Self-consistency requires us to find complex coefficients $a_j$ such that the
time-independent term in eq.~\eqref{eq:interference} recovers the 
static background density $\rho_\text{BG}(\bm x)$, i.e.:
\begin{equation}
    \label{eq:self-consistency}
    \rho_\text{BG}(\bm x) \stackrel{!}{=} \sum_j |a_j|^2|\psi_j(\bm x)|^2 =
    \left\langle|\psi|^2\right\rangle \;.
\end{equation}
A few remarks are in order. Firstly, eq.~\eqref{eq:self-consistency} only fixes
the coefficient moduli, but leaves the phases $\omega_j$ unspecified. We follow \cite{Yavetz2022} and sample $\omega_j \sim U\left([0,2\pi)\right)$. $\langle|\psi|^2\rangle$ thus denotes the expectation value over all phases.

Secondly, a direct comparison between solution of eqs.~\eqref{eq:se}-\eqref{eq:poisson} and eqs.~\eqref{eq:nlse}-\eqref{eq:fdm_poisson} for virialised halos show satisfying
agreement \citep{Yavetz2022}.

Thirdly, although the requirement
eq.~\eqref{eq:self-consistency} implies that each eigenstate must obey the same symmetries as the steady-state background --- spherical for haloes, cylindrical for filaments ---$|\psi(\bm x, t)|^2$, and in particular the interference term in eq.~\eqref{eq:interference}, does not. As shown in Fig.~\ref{fig:showcase_crosssection}, interference fringe features beyond the
assumed symmetry in $\rho_\text{BG}$ can thus be realised.

Finally, despite our focus on FDM interference, there is no \emph{a priori}
reason to restrict ourselves to FDM-only steady-state background models.
As shown in \cite{Yavetz2022}, as long as the set of eigenstates is able to
reproduce all properties of $\rho_\mathrm{BG}$ on the spatial scales of interest
(in this work $>\SI{1}{\kilo\parsec}$), 
the latter is a viable input to
the interference model. This opens up the possibility to use the above model
prescription as a post-processing tool for CDM simulations of steady-state
objects, an intriguing application we leave to a future work.

To apply the outlined procedure to the case of FDM filaments, three ingredients
are required: (i) a background density model (ii) a library of eigenstates
and eigenenergies and (iii) a procedure to satisfy eq.~\eqref{eq:self-consistency}.
We address each point in Sec.~\ref{sub:filament_cylinder}-\ref{sub:reconstruction}.

\subsection{\label{sub:filament_cylinder} FDM filaments as
infinitely long, isothermal cylinders}
Let us map the general discussion of Sec.~\ref{sub:interference} to the
special case of fuzzy filaments with the goal to establish eqs.~\eqref{eq:se} - \eqref{eq:poisson} as a viable approximation to eqs.~\eqref{eq:nlse} - \eqref{eq:fdm_poisson}. 

From the onset, we stress that there is no a priori reason to
expect that the entire cosmic filament population may be attributed with
universal properties like a complete relaxation into virial equilibrium or a
NFW-profile analogue. 
Even from an idealised theoretical viewpoint \citep{Zeldovich1970, Bond1996}, filaments 
represent only partially collapsed objects, still expanding or contracting along
their spines on a time scale comparable to the Hubble time, $t_\text{H}$, ultimately
turning into fully collapsed, virialised haloes. This is no different for
FDM structure formation and an enlightening volume rendering of this process is
shown in \cite{Nori2021}.

To complicate matters,
simulations suggest that a variety of factors may determine the physical conditions
and morphology of a filament, most notably spatial scale
\citep[e.g.][]{Colberg2005}, cosmic environment \citep{Ramsoey2021}, and properties
of DM \citep{Mocz2019, May2022}. 
How do we arrive at a time-independent, linear approximation of eq.~\eqref{eq:nlse} in light of this physically diverse filament population?

In what follows, we distinguish between two limiting cases depending on the
kinematic state:
\emph{bulk-flow dominated} and \emph{quasi-virialised} filaments.

Bulk-flow dominated filaments \citep[e.g.][]{Odekon2022} represent
$>\mathcal{O}(\SI{1}{\mega\parsec})$
structures acting as accretion channels for DM and gas
into the gravitational well of the adjacent DM haloes. Their radial
density structure is consistent with a broken power-law profile with a power law index
$\gamma = -2$ in the outskirts \citep{Colberg2005,Dolag2006,Aragon-Calvo2010,Zhu2021}
and core regions consistent with $-1\leq \gamma \leq 0$. 
It is clear that eqs.~\eqref{eq:se}-\eqref{eq:poisson} can not be a \emph{bona fide} description of
this filament population as the impact of the environment, in particular inflow and outflow of matter 
is excluded: $\psi$ and its steady-state density $\rho_\text{BG}$ model an
isolated, self-gravitating object. We return to the question of how one may
model bulk-flow dominated filaments in Sec.~\ref{sec:conclusion}.

The quasi-virialised case \citep{Stodolkiewicz1963, Ostriker1964}, by contrast, 
constitutes a theoretically idealised
scenario in which filaments are approximated locally by an infinitely long,
isothermal, cylinder. In it each
cross section attained a steady-state, virial equilibrium and dynamics along the
longitudinal direction are suppressed. The model is appealing as (i) it maps
well to the assumptions leading to eqs.~\eqref{eq:se} - \eqref{eq:poisson} and
(ii) the derivation of the steady-state real-space density $\rho_\mathrm{BG}$ and
even its phase space distribution is tractable. We clarify these aspects
in Secs.~\ref{subsub:real_space} - \ref{subsub:phase_space}.

Interestingly, recent high-resolution simulations of intermediate-size
($\lesssim\mathcal{O}(1)\si{\mega\parsec}$) CDM filaments around Milky Way-like
progenitors \citep{Ramsoey2021} are able to reproduce filaments of this type
--- a statement found to be true across redshifts $z=3-8$. Moreover, \cite{Eisenstein1997} propose to estimate the dynamical mass of \emph{large} scale filaments under isothermal cylinder conditions and find the approximation to work reasonably well when compared against $N$-body simulations. 

The remainder of this work focuses on the wave function reconstruction in the
quasi-virial limit as well as observable changes in statistical measures due to
the \emph{presence} of interference sourced by such filaments, cf. Sec.~\ref{sec:statistics}.
Dynamical implications of interference \citep{Dalal2022} will not be considered.
This allows us to construct the interference field in eq.~\eqref{eq:interference} at 
only one instance of cosmic time $t$ and thus at a frozen scale factor value $a=a(t)$.
We choose $z=4$ to be (i) within the applicability region of the isothermal
cylinder approximation \citep{Ramsoey2021}, (ii) near the regime in which the
large scale behavior of the FDM power spectrum is dominantly determined by the
choice of initial conditions \citep{May2022} 
(thereby simplifying the construction of the filament mass
function in Sec.~\ref{sub:mass_function}), and (iii) compatible with the results
of of \cite{Gough2024}, with which we make contact
in Sec. \ref{sec:discussion}.

\subsubsection{\label{subsub:real_space}Real Space Density}

We first argue that the interference
contained in the full wave function has negligible impact on the gravitational
potential.
For this, we posit that filaments may be approximated as isolated, infinitely long,
at this stage not necessarily isothermal, cylinders.
Consequently, the right hand side of eq.~\eqref{eq:fdm_poisson} reduces to
$|\psi|^2$, which is, by assumption, now constant along the filament spine. The gravitational
potential for a non-axially symmetric source $|\psi(\bm x)|^2 = |\psi(R, \Phi)|^2$ reads:
\begin{equation}
    \label{eq:poisson_cylinder}
    V(\bm x) 
    = \int_{\mathbb{R}^2} \text{d}^2x' G(\bm x, \bm x')\left(\frac{4 \pi
    G}{a}|\psi(\bm x')|^2\right),
\end{equation}
with $G(\bm x, \bm x') = (2\pi)^{-1} \log(|\bm x - \bm x'|)$ denoting the
free-space Greens function in polar coordinates. Compared to a canonical
$1/r$-kernel in spherical symmetry, a $\log$-potential shows strong smoothing
properties when used in the convolution of eq.~\eqref{eq:poisson_cylinder},
effectively rendering the interference contribution in $|\psi|^2$ to the total
gravitational potential subdominant.

\begin{figure}
    \centering
    \includegraphics[width=\columnwidth]{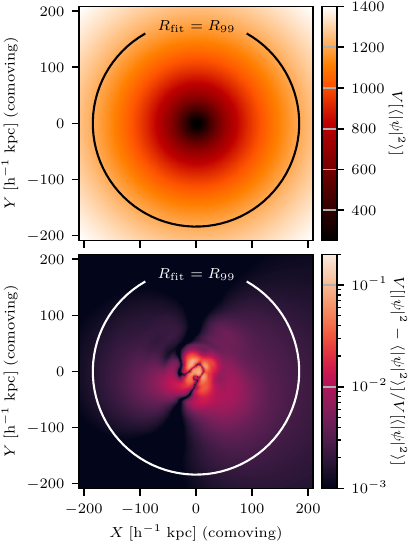}
    \caption{\label{fig:potential_smoothing}
        \textbf{(Top)}
        Gravitational potential sourced by the
        axial-symmetric contribution $\langle |\psi|^2\rangle_\Phi = \langle
        |\psi|^2\rangle$ of the wave function depicted in Fig.~\ref{fig:showcase_crosssection}. 
        \textbf{(Bottom)} 
        Relative importance of the potential perturbation sourced by the interference term
        alone. Neglecting the latter in the computation of the
        total potential in eq.~\eqref{eq:fdm_poisson} only incurs $\mathcal{O}(1-10\%)$ errors.
    }
\end{figure}

Let us illustrate this smoothing effect based on the strongly oscillating
wave function shown in Fig.~\ref{fig:showcase_crosssection}. 
For this, we compare the potential produced by the non-axially symmetric,
interference-modulated density $|\psi|^2$, with the axial-symmetric
interference-free contribution $\langle|\psi|^2\rangle$.
For each contribution we compute the gravitational potential according to eq.~\eqref{eq:poisson_cylinder} \citep{Hejlesen2019}
and compare their relative importance in Fig.~\ref{fig:potential_smoothing}. 

Evidently, the interference
part of the wave function induces small, mostly per-cent level, changes to the total 
gravitational potential sourced by $|\psi|^2$. Thus, we may substitute $|\psi|^2
\to \langle |\psi|^2\rangle$ in eq.~\eqref{eq:poisson_cylinder} and only incur a
relatively small error in the potential.

Next, we seek for an axial-symmetric filament background density
$\rho_\text{BG}(R)$ that may replace the wave function all together and thus
removes the nonlinear coupling between Schr\"odinger's and Poisson's equation.

On scales larger than $\lambda_\text{dB}$ the Schr\"odinger-Vlassov
correspondence \citep{Uhlemann2014, Mocz2018} allows us to treat FDM as
collisionless particles.
In accordance with our introductory remarks, let the behaviour of these particles be described by
a velocity dispersion tensor of the form:
\begin{equation*}
    \bm \sigma^2
    =\mathrm{Diag}\left(\langle v_r^2 \rangle, \langle v_\phi^2\rangle,
    0\right)\;,
\end{equation*}
and thus neglecting longitudinal dynamics,
$\langle v_z \rangle = \langle v_z^2\rangle = 0$, and bulk
rotation, $\langle v_\phi\rangle = 0$.
Isothermal conditions are imposed by assuming the radial moment, $\langle
v_r^2\rangle$, azimuthal moment, $\langle v_\phi^2 \rangle$, and consequently
the anisotropy parameter,
\begin{equation}
    \beta \equiv 1 - \frac{\langle v_\phi^2 \rangle}{\langle v_r^2 \rangle}
    \;,
\end{equation}
are constant throughout the filament.

Starting from the cylindrical Jeans equation \citep{Binney2008} and above
dispersion tensor, 
\citet{Eisenstein1997} derive closed form, analytic solutions
for the density $\rho$ and its associated line density $\mu(r)$:
\footnote{
    This is a generalisation of the isotropic, $\beta = 0$, case
    considered in \cite{Stodolkiewicz1963, Ostriker1964}.
}
\begin{align}
    \label{eq:steady_state_density}
    \rho(r\mid r_0, \sigma^2, \beta) &= \frac{(2-\beta)^2\sigma^2}{2\pi G
    r_0^2}\frac{y^{-\beta}}{\left(y^{2-\beta} + 1\right)^2}\;, \quad
    y\equiv\frac{r}{r_0}\;,\\
    \label{eq:steady_state_mass_density}
    \mu(r\mid r_0, \sigma^2, \beta) &\equiv 2 \pi \int_0^r\text{d}r' r' \rho(r') 
    = \frac{(2-\beta)\sigma^2}{G}\frac{y^{2-\beta}}{y^{2-\beta} + 1} 
    \;.
\end{align}
$r_0$ denotes the transition scale between the
small radii, $\rho \sim r^{-\beta}$, and large radii behaviour, $\rho \sim r^{\beta-4}$. 
In contrast to an NFW profile or the isothermal sphere, the integrated (line) mass 
$\mu$ remains finite as $r\to\infty$:
\begin{equation}
    \label{eq:line_mass_dispersion}
    \mu \equiv \frac{(2-\beta)\sigma^2}{G} \;.
\end{equation}
We call filaments defined as above ``quasi-virial'' since the relation between total line mass $\mu$, dispersion
$\sigma^2$ and anisotropy parameter $\beta$ is a direct consequence of the tensor virial theorem 
applied to the infinite cylinder geometry. The isotropic, $\beta=0$, derivation is given in
\cite{Fiege2000}.

As noted earlier, \cite{Ramsoey2021} find intermediate-sized CDM filaments in
Milky Way sized progenitor haloes to adhere to the
isotropic profile in eq.~\eqref{eq:steady_state_density}.\footnote{
    Embedding the cylinder into the isothermal solution of a steady-state sheet
    improves the correspondence with the simulated density profiles even further.
    Since our focus is on filaments only, we neglect this contribution.
}
More precisely, their filament sample is consistent with a quasi-static
evolution of eq.~\eqref{eq:steady_state_density} between redshift $z=3-8$, i.e.
the functional form of the density profile remains valid throughout this redshift bin and only a 
time dependence in the velocity dispersion $\sigma^2\to\sigma^2(a)$ and scale radius $r_0\to r_0(a)$ 
is introduced. For reference, at redshift $z_*=4$ one finds $r_0(a_*) \simeq \SI{10}{kpc}$ and
$\sqrt{\sigma^2}(a_*) \simeq \SI{10}{\kilo\meter\s^{-1}}$ \citep{Ramsoey2021}.

It is interesting to note that, although the forgoing discussion assumed
collisionless dynamics, by virtue of the large-scale equivalence of CDM and FDM, a stability analysis \citep{Desjacques2018} suggests that eq.~\eqref{eq:steady_state_density} is a stable density configuration for SP.\footnote{
    Adding an attractive, local interaction of the form $\lambda |\psi|^2\psi$
    introduces a critical line mass $\mu_c$ above which even infinitely long
    cylinders are unstable and collapse radially.
}

\cite{May2022} extract radial filament density profiles for CDM and FDM
from a comparable set of cosmological simulations and find \emph{both} populations to have cuspy
inner densities at $z=3$. 
We may realise such cuspy profiles by assuming a radially biased velocity
anisotropy, i.e. $\beta > 0$.

We thus arrive at the sought-after, axial-symmetric, comoving background density
$\rho_\mathrm{BG}$:
\begin{align}
    \label{eq:rho_bg}
    \rho_\text{BG}(R, a) &= \rho\left(Ra\mid r_0(a), \sigma^2(a), \beta\right)
    a^3 \;.
\end{align}
It is this density that sources the potential within which the filament wave
function evolves.

\subsubsection{\label{subsub:phase_space}Phase Space Distribution Function}

With a background density, eq.~\eqref{eq:rho_bg}, in place we
ask what the associated phase space distribution function (DF) looks like. Constructing
the latter for generic axial-symmetric systems, e.g. galactic disks, is a daunting task, 
both theoretically \citep{Hunter1993} and numerically \citep{Petac2019} --- especially in the 
case of prolate objects \citep{Merritt1996} like \emph{finite} extend filaments.

Fortunately, in the case of \emph{infinitely} long cylinders, c.f. Sec.~\ref{subsub:real_space}, the situation simplifies drastically and is
conceptionally close to the situation found for spherically symmetric but
anisotropic systems \citep{Binney2008}.

According to Jeans' theorem, we may assume that a steady-state DF is
a function of three isolating integrals of motion only. In the case of a
classical particle moving in the gravitational potential sourced by an
infinitely long cylinder, these are trivially identified as the specific energy contribution
of both the longitudinal, $E_z$, and transversal motion, $E$, as well as the specific
angular momentum $L$:
\begin{equation}
    E = \frac{1}{2a^2} (u_R^2 + u_\Phi^2) + V(R)\;,\quad 
    E_Z = \frac{u_Z^2}{2a^2}\;,\quad 
    L=R u_\Phi \;.
\end{equation}
where $u_i \equiv p_i/m$ denote the velocities associated with the
momenta conjugate to comoving coordinates, i.e. $p_i = \frac{\partial
L}{\partial x_i}$ and L as effective Lagrange function of a classical particle
in an expanding universe \citep{Peebles1980, Bartelmann2015}.
We remind the reader that the scale factor is fixed and energy is therefore a conserved quantity.

Jeans' theorem then suggests the DF ansatz:\footnote{
    Notice that we construct a DF that depends on the 
    \emph{magnitude} of the specific angular momentum $|L|$ rather
    than its conserved value $L=R u_\Phi$ --- a
    choice we make to arrive at eq.~\eqref{eq:inverse_problem} but also to
    remain consistent with the discussion of Sec.~\ref{subsub:real_space} where
    we excluded the possibility of a bulk rotation.
}
\begin{equation}
\begin{split}
    \label{eq:df_ansatz}
    \tilde f(E, E_Z, L) &= \mathcal{N} f(E, |L|) \delta_D\left(\frac{u_Z^2}{2 a^2}
    - E_Z\right)\quad \text{with}\\ 
        \mathcal{N} &\equiv \mu
        \left(
            \int\dd R\dd \Phi\dd^3 \bm u R \tilde f(E, E_Z, |L|)
        \right)^{-1},
\end{split}
\end{equation}
and $\mu$ as defined in eq. \eqref{eq:line_mass_dispersion}.
Marginalising over velocity space provides us with the radial density
implied by the DF of eq.~\eqref{eq:df_ansatz}:
\begin{equation}
    \begin{split}
        \label{eq:df_rho}
        \rho_\text{DF}(R) &= \iiint \dd u_R \dd u_\Phi \dd u_Z \tilde f(E, E_z, |L|) \\
                &= 4a^2 \int\limits_{V(R)}^{\infty} \dd E  \hspace{-1.5em}
                \int\limits_{0}^{\sqrt{2a^2(E-V(R))}}\hspace{-1.5em}\dd u_\Phi
                \frac{f(E, |L|)}{\sqrt{2a^2(E-V(R)) - u_\Phi^2}} \;.
    \end{split}
\end{equation}

We wish to invert this expression to obtain $f(E, |L|)$. However, eq.~\eqref{eq:df_rho}
is ill-defined without further restrictions on the angular momentum dependence
of $f$ as different radial velocity dispersions can in principle realise the
same radial density profile. In accordance with Sec.~\ref{subsub:real_space}, we break this
degeneracy by enforcing a constant anisotropy $\beta$. 
Either through direct computation or by application of the augmented density
formalism \citep{Dejonghe1986}, we find:
\begin{equation}
    \label{eq:constant_beta_df}
    f(E, |L|) = |L|^{-\beta} g(E)
\end{equation}
to guarantee $\beta = \text{const.}$. The yet to be determined function $g(E)$
then follows by demanding that marginalising eq.~\eqref{eq:constant_beta_df}
recovers the background density of eq.~\eqref{eq:rho_bg}. We set $\rho_\text{DF}(R) =
\rho_\text{BG}(R)$ in eq.~\eqref{eq:df_rho} and perform the $u_\Phi$-integration 
to arrive at:
\begin{equation}
    \label{eq:inverse_problem}
    \rho_\text{BG}(R) = 
    2 a^2\sqrt{\pi}
    \frac{\Gamma\left(\frac{1-\beta}{2}\right)}{\Gamma\left(\frac{2-\beta}{2}\right)}
    R^{-\beta}
    \int\limits_{V(R)}^{\infty} \dd E
    \frac{g(E)}{\left[2a^2(E-V(R))\right]^{\beta/2}}\;,
\end{equation}
with $\beta \in (-\infty, 1)$. For the physically relevant range of anisotropies $\beta \in [0,1)$, eq.~\eqref{eq:inverse_problem} is equivalent to an Abel integral equation and thus
existence of a solution for $g(E)$ is guaranteed.\footnote{
    To solve eq.~\eqref{eq:inverse_problem} in practice, we approximate $\log g(E)$
    as a multilayer perceptron \citep{Hastie2009}, perform the
    numerical integration as forward pass and find the perceptron
    weights by minimising the least square difference to $\rho_\text{BG}$.
}

This concludes the construction of the background model. 
In summary, in the limit of quasi-virial filaments, it is possible to find a
self-consistent description 
in real and phase space. We use $\rho_\text{BG}$ in eq.~\eqref{eq:rho_bg} to find the gravitational
potential $V$ via eq.~\eqref{eq:poisson_radial} and use the density-potential
pair as
input to the inverse problem of eq.~\eqref{eq:inverse_problem}, the solution of
which gives us access to the DF.

\begin{figure}
    \centering
    \includegraphics[width=\columnwidth]{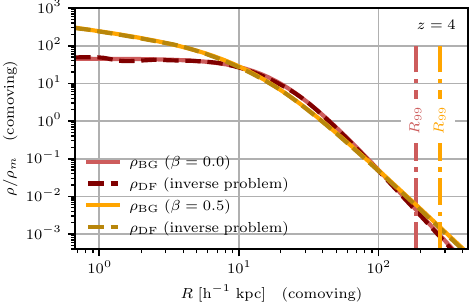}
    \caption{\label{fig:rho_bg_df}
        Radial filament densities obtained from eq.~\eqref{eq:rho_bg} (solid
        lines) and the solution of the inverse problem of eq.~
        \eqref{eq:inverse_problem} (dashed lines), i.e. the density implied 
        by the DF. Vertical lines depict the fit radius $R_{99}$ within which
        $99\%$ of the total line mass $\mu$ are contained.
    }
\end{figure}

\begin{figure}
    \centering
    \includegraphics[width=\columnwidth]{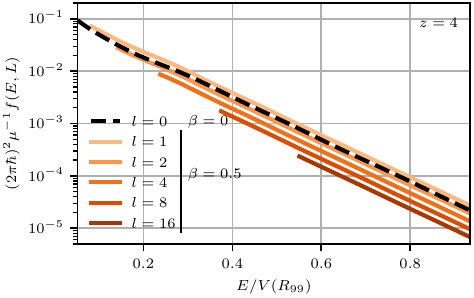}
    \caption{\label{fig:df}%
        Recovered DFs for the densities depicted in Fig.~\ref{fig:rho_bg_df} for
        various values of the specific angular momentum $L=l\times\hbar m^{-1}$. Note that (i) we
        only show $L=0$ for $\beta=0$ as the DF
        is uniform in $L$ for isotropic conditions, cf. eq.~
        \eqref{eq:constant_beta_df}, (ii) energies are bound from below
        by $E_c(L)$, the energy of a circular orbit defined as
        the minimum of the effective potential $V(R) + L^2/2a^2R^2$ and (iii)
        both DFs recover a Boltzmann-like, exponential behaviour in $E$. We
        emphasize that this functional behaviour is not imposed by our inversion
        of eq.~\eqref{eq:inverse_problem}.
    }
\end{figure}

We refer to Fig.~\ref{fig:rho_bg_df} for an illustration of the density, both from eq.~\eqref{eq:rho_bg} and eq.~\eqref{eq:inverse_problem} for two values of the anisotropy parameter $\beta$.
The correspondence between both profiles is satisfying, and therefore
demonstrates the effectiveness of our numerical inversion for eq.~\eqref{eq:inverse_problem}, the solution of which is depicted in Fig.~\ref{fig:df}. For both choices of $\beta$, the inversion converges to a DF that
is exponential in the specific energy $E$.


\subsubsection{\label{subsub:library}Eigenstate Library}

We may realise an infinitely long, cylindrical geometry by solving eqs.~\eqref{eq:se}-\eqref{eq:poisson} in a domain unbound in $X,Y$-direction and periodic in $Z$.
A factorisation of the eigenmode $\psi_j(\bm x)$ into:
\begin{equation}
    \label{eq:factorisation}
    \psi_j(\bm x) = \psi_{nl}(R,\Phi) =
    \sqrt{\frac{\mu}{2\pi}}R_{nl}(R)e^{il\Phi} \;, \;
    \int_0^\infty \dd R\, R\, R_{nl}^2(R) = 1,
\end{equation}
reduces eqs.~\eqref{eq:se}-\eqref{eq:poisson} to:
\begin{subequations}
\begin{align}
    \begin{split}
    \label{eq:radial_states}
    E_{nl} u_{nl}
    &= -\frac{\hbar^2}{2m^2a^2}\partial^2_R u_{nl} + V_{\mathrm{eff},
        \psi}
    u_{nl}\;, \\
    V_{\mathrm{eff},\psi}(R) &= V(R) + \frac{\hbar^2}{2m^2a^2}\frac{l^2 - 1/4}{R^2} \;,
    \end{split}\\
    \begin{split}
    \label{eq:poisson_radial}
    \frac{1}{R}\frac{\partial}{\partial R}\left(R\frac{\partial}{\partial R} V\right) 
    &= \frac{4\pi G}{a}\rho_\mathrm{BG}(R)
    \end{split}\,,
\end{align}
\end{subequations}
with radial eigenstates $R_{nl}(R) = \sqrt{R} u_{nl}(R)$ and $n \in \mathbb{N}, l \in \mathbb{Z}$
as radial (number of nodes of $R_{nl}$) and angular quantum number respectively.
The Hamiltonian is symmetric in $l$. Consequently, the specific energy
eigenvalues satisfy $E_{n,-l} = E_{nl}$ as well as
$R_{n,-l}(R) = R_{nl}(R)$ and we may restrict the construction of the library to
positive angular momenta.

The numerical treatment of eq.~\eqref{eq:radial_states} is more challenging than
the spherically symmetric halo case, in part because of the form the angular
momentum barrier takes, see Appendix~\ref{appendix:numerics} for a discussion. 

\begin{figure*}
    \centering
    \includegraphics[width=\textwidth]{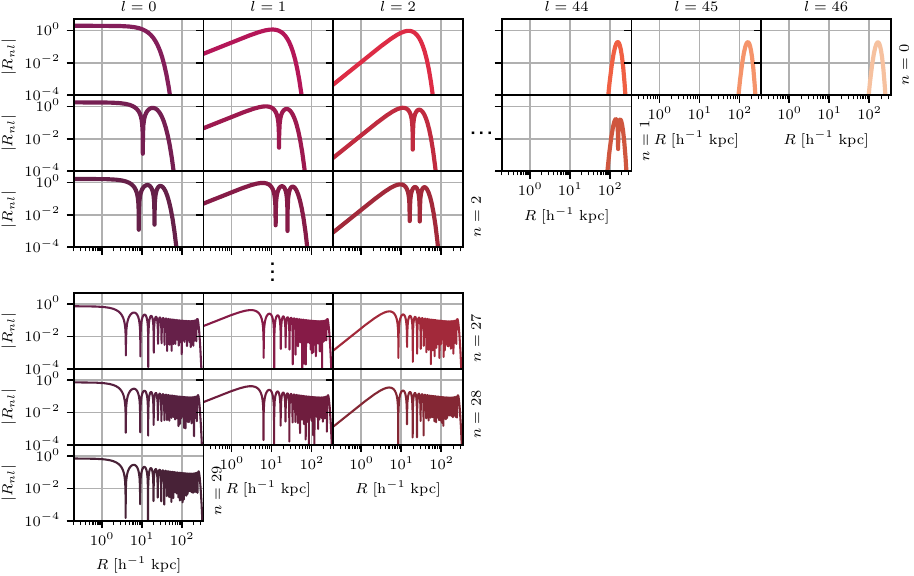}
    \raisebox{4.6cm}[0pt][0pt]{%
        \hspace{9.7cm}%
        \parbox{8cm}{
            \caption{\label{fig:eigenstates_overview}%
                Subset of the eigenstate library for $m=\SI{2e-22}{\eV}\;,\;\beta=0$ at
                redshift $z=4$. The total library for this parameter combination
                contains $J=746$ states. As is the case for halos,
                the small radii behaviour of radial eigenstates is determined by the
                angular momentum $R_{nl}(R) \sim R^l$, while $n\in\mathbb{N}$ is
                equivalent to the number of nodes in $R_{nl}$. For fixed $l$,
                the maximal $n$ is set by $E_{nl} < V(R_{99})$ and
                $l_\mathrm{max}$ follows from $\min\left(V_{\mathrm{eff},\psi}\right) < V(R_{99})$
            }
        }
    }
\end{figure*}

For a fixed value of $l$, the radial eigenstates of eq.~\eqref{eq:radial_states}
are found by a Chebyshev pseudospectral discretisation that respects the parity
properties of the modes $\psi_j(\bm x)$ in eq.~\eqref{eq:factorisation} 
\citep{Lewis1990, Trefethen2000}.
This results in a dense matrix representation of the Hamiltonian, $\bm H_l$, for which we
compute the eigensystem and retain all modes with
$E_{nl} \leq V(R_{99})$. $R_{99}$ corresponds to the radius within
which $99\%$ of the total line mass $\mu_\text{BG}$ are contained and the stated
inequality discards all eigenstates with a classically allowed region larger
than our fiducial fit radius $R_\text{fit} = R_{99}$. 
The procedure is repeated for increasing values of $l$ until the minimum of the
effective potential, $V_{\mathrm{eff},\psi}$, surpasses our cutoff energy $V(R_{99})$.
Note that this procedure does not set the number of eigenstates, $J_l$, a priori, but finds \emph{all} modes up
to the cutoff energy as long as the spatial discretization satisfies $\text{rank}(\bm H_l) > J_l$.
We depict a library excerpt for $m=\SI{2e-22}{\eV}$
and an isotropic background density at redshift $z=4$ in Fig.~\ref{fig:eigenstates_overview}.

Depending on input parameters
$\{z, m, \beta\}$ a complete eigenstate library contains
$\mathcal{O}(10^2)-\mathcal{O}(10^4)$ modes (and consequently the same number of mode
coefficients $a_j \in \mathbb{C}$). Thus, we end up with a highly flexible model
potentially prone to overfitting. Sec.~\ref{sub:reconstruction} outlines how the DF constructed in
Sec.~\ref{subsub:phase_space} may be used to arrive at a minimal model that still recovers the 
steady-state density $\rho_\mathrm{BG}$ as closely as possible.

\subsection{\label{sub:reconstruction}Wave Function Reconstruction}
Under cylindrical symmetry and ansatz eq.~\eqref{eq:radial_states}, 
the self-consistency condition, eq.~\eqref{eq:self-consistency}, is recast to:
\begin{equation}
    \label{eq:self-consistency_filament}
    \rho_\mathrm{BG}(R)  = \frac{\mu}{2\pi} \sum_{n,l \ge  0}
    N_{l}|a_{nl}|^2|R_{nl}(R)|^2\;, \;
    N_l = \begin{cases}
        1, &l=0\\
        2, &l>0
    \end{cases} \;.
\end{equation}

There exist multiple approaches for computing the coefficients $a_{nl}$.
\citet{Widrow1993} propose a physically-intuitive procedure and set the
coefficients according to the DF --- after all $|a_{nl}|^2$ may be interpreted as the
probability of finding state $|\psi_{nl}|^2$ and the value of the DF represents
the classical probability of a system state with energy $E_{nl}$ and angular
momentum $L_{nl}$. In the cylindrical case considered here it is therefore natural to assume:
\begin{equation}
    \label{eq:kaiser_coefs}
    |a_{nl}| \propto f(E_{nl}, L_{nl}) \Delta E \Delta L\;.
\end{equation}

\citet{Lin2018, Dalal2021} show the effectiveness of this approach in the case of virialised, fully isotropic haloes for which the DF is a function of energy only. More precisely, eq.~\eqref{eq:kaiser_coefs} produces FDM haloes closely
following an NFW halo upon time or phase averaging. However, deviations
are found at small radii. An explanation for this behaviour is given by
\cite{Yavetz2022}: eq.~\eqref{eq:kaiser_coefs} is only correct in the WKB-limit of
high energies $E\gg V$. Hence, only these modes are assigned correct
coefficients. Since highly excited eigenstates
have broader spatial extent, cf. Fig.~\ref{fig:eigenstates_overview}, 
they dominate the mode composition in the halo outskirts and produce the correct
density profile. Low energy eigenstates, on the other hand, don't contribute to the
outskirt density and their poorly chosen coefficients only contribute
to deviations in the core region.

Appendix~\ref{appendix:wkb} derives the high energy WKB correspondence between DF and the
wave function coefficients for the case of infinitely long filaments of line mass
$\mu$. One finds:
\begin{equation}
    \label{eq:wkb}
    |a_{nl}|^2 \sim |a_{nl,\mathrm{WKB}}|^2\equiv\frac{(2\pi\hbar)^2}{m^2\mu}
    f(E_{nl}, L_{nl}) \;,\; L_{nl} = \max(L_0, l) \frac{\hbar}{m}\;,
\end{equation}
where $L_{nl}$ represents an effective mapping from quantum number $l$ to
the classical, specific angular momentum $L$. Its form is motivated by Langer's
correction \citep{Langer1937} and modified at $l=0$ where anisotropic DFs diverge,
cf. eq \eqref{eq:constant_beta_df}. We fix the constant $L_0 < 1$ such that a
wave function with coefficients set according to eq.~\eqref{eq:wkb} reproduces the
background density $\rho_\mathrm{BG}$ up to $R \sim \SI{1}{\kilo\parsec}$. 

\begin{figure}
    \centering
    \includegraphics[width=\columnwidth]{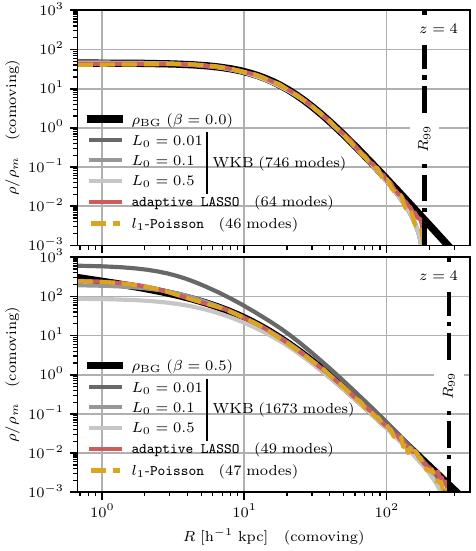}
    \caption{\label{fig:density}%
        Comparison of various coefficient estimators $|a_{nl}|^2$ for 
        $m=\SI{2e-22}{\electronvolt}$ with $\beta=0$ (top) and $\beta=0.5$ (bottom)
        at $z=4$. While the WKB result, eq.~\eqref{eq:wkb}, reproduces the
        background density well without any optimisation, the posterior modes
        \texttt{adaptive LASSO} and $l_1$-\texttt{Poisson} show identical
        accuracy while reducing the number of involved modes significantly.
    }
\end{figure}

For this we turn to Fig.~\ref{fig:density} illustrating
the effectiveness of the WKB assignment scheme in the case of quasi-virialised
filaments. Under isotropic conditions (top panel) the DF is independent of
$L$ rendering the choice of $L_0$ irrelevant. For $\beta > 0$ (lower panel),
an ad-hoc value of $L_0 = 0.1$ recovers the cuspy density profile reasonably
well up to $R \sim \SI{1}{\kilo\parsec}$.

In general, the WKB result yields a filament wave function in satisfying agreement
with the input density $\rho_\mathrm{BG}$ --- without having to optimise the
coefficients in any way. We emphasise that the results of Fig.~\ref{fig:density} are also a
non-trivial convergence test of both the DF construction (Sec.~\ref{subsub:phase_space}) and the eigenstate library (Sec.~\ref{subsub:library})
which are independent components of our model and only related by eq.~\eqref{eq:wkb}, yet able to recover $\rho_\text{BG}$ in practice.

\cite{Yavetz2022} propose a regression approach to eq.~\eqref{eq:self-consistency_filament} by setting
\begin{equation}
    \label{eq:regression}
    |a_{nl}|^2 = 
    \argmax_{|a_{nl}|^2} \mathcal{L}\left(\rho_\mathrm{BG} \mid \langle
    |\psi|^2\rangle(|a_{nl}|^2)\right)\;,
\end{equation}
with a gaussian likelihood $\mathcal{L}$ that allows to reconstruct FDM haloes which are consistent with time averaged density profiles derived from fully-fledged simulations of eqs.~\eqref{eq:nlse}-\eqref{eq:fdm_poisson}.

Note that given the size of an initial eigenstate library with
$\mathcal{O}(10^2)-\mathcal{O}(10^4)$ coefficients to fit, 
the convergence of eq.~\eqref{eq:regression} and overfitting are natural
concerns. In fact, \cite{Yavetz2022} report the need of an iterative fitting
approach to arrive at density profiles resembling the time static input density.
They combat this slow convergence by manually imposing isotropy. By
binning the energy spectrum, modes that fall into the same energy bin are forced
to have identical expansion coefficients.

Here we propose a physically-informed regularisation of the regression approach
of eq.~\eqref{eq:regression} with the WKB result in eq.~\eqref{eq:wkb}. This is
achieved by (i) putting an exponential prior, consistent with the form of the DF
noted above, on each coefficient
$|a_{nl}|^2$ and (ii) setting the sought-after coefficients as the mode of 
the posterior. More precisely, we use:
\begin{align}
    \label{eq:our_approach}
    |a_{nl}|^2 &= 
    \argmax_{|a_{nl}|^2} 
    \left\{
        \mathcal{L}\left(\rho_\mathrm{BG} \mid \langle
        |\psi|^2\rangle\right) \prod_{nl} p\left(|a_{nl}|^2 \mid |a_{nl,\mathrm{WKB}}|^{2}\right)
    \right\}\;,
\end{align}
and set:
\begin{equation}
    \label{eq:prior}
    p\left(|a_{nl}|^2 \mid |a_{nl, \mathrm{WKB}}|^{2}\right) = \begin{cases}
        \frac{1}{|a_{nl, \mathrm{WKB}}|^{2}} \exp\left(-\frac{|a_{nl}|^2}{|a_{nl,
        \mathrm{WKB}}|^{2}}\right) & |a_{nl}|^2 > 0  \\
        0 & \text{else}
    \end{cases}\;.
\end{equation}
With this choice the prior expectation is
$\mathbb{E}[|a_{nl}|^2]~=~|a_{nl,\mathrm{WKB}}|^{2}$ and it is in this sense
that we encode the asymptotic result of eq.~\eqref{eq:wkb} in the regression
approach.

Our choice for the likelihood is determined by the process that generates the noisy
realisation of $\rho_\mathrm{BG}$ against which we fit. The canonical choice is a Gaussian error
model of spatially uniform variance $\Sigma^2$, i.e. $\log \rho_\mathrm{BG}(R_i)
\sim \mathcal{N}(\log|\psi_\mathrm{BG}(R_i)|^2 \mid \Sigma^2)$:
\begin{equation}
    \mathcal{L}\left(\rho_\mathrm{BG} \mid \langle |\psi|^2\rangle\right)
    = \prod_i\mathcal{N}\left(\log |\psi_\mathrm{BG}(R_i)|^2 \mid \log \langle
    |\psi(R_i)|^2\rangle, \Sigma^2\right)\,.
\end{equation}
Equation~\eqref{eq:our_approach} is then identical to an adaptive LASSO estimator \citep{Zou2006}, 
but with the per-coefficient regularisation strength fixed by the value of the DF.
This estimator is most suited for scenarios in which fuzzy filaments are generated ex situ, as is the case in this work. The error variance $\Sigma^2$ then constitutes a user-defined hyperparameter.

More suited for post-processing applications of simulated, interference-free DM filaments is a process that captures the intrinsic noise of the
particle ensemble enclosed in the filament. This can be done via a spatial
Poisson process of variable density. Let $y_i$ be the
number of fiducial DM particles in an annulus between $R_i$ and $R_{i+1}$ around the 
filament spine. Assuming Poissonian statistics, i.e. $y_i \sim \mathrm{Poisson}(M_{ij}|a_j|^2)$, the likelihood 
takes the form:
\begin{align}
    \label{eq:poisson_process}
    \mathcal{L}\left(\rho_\mathrm{BG} \mid \langle |\psi|^2\rangle\right)
    &= \prod_i
    \frac{(M_{ij}|a_j|^2)^{y_i}}{y_i!} \exp\left(-M_{ij}|a_j|^2\right), \\
    M_{ij} &= 2\pi\int_{R_i}^{R_{i+1}}\text{d}R\, R |\psi_j|^2 \;.
\end{align}

Irrespective which likelihood is used, the WKB prior \eqref{eq:prior}
introduces $l_1$-regularisation into the optimisation problem. In combination with the
maximum-a-posterior value,
eq.~\eqref{eq:our_approach} enjoys the desirable feature selection property: 
it promotes sparsity in $|a_{nl}|^2$ so that modes which do not contribute to the
reconstruction of $\rho_\mathrm{BG}$ are driven to zero \emph{exactly}.
We are thus able to construct a minimal model that reproduces the
steady-state background. Combining this approach with special-purpose optimisation
algorithms \citep{Harmany2012,Parikh2014} allows for fast convergence without the need 
of an artificial energy binning.

The effectiveness of eq.~\eqref{eq:our_approach} is showcased in Fig.~\ref{fig:density}.
One finds the posterior modes \texttt{adaptive LASSO} and $l_1$-\texttt{Poisson}
to be as accurate as the WKB assignment and simultaneously reduce the size of the 
eigenstate library significantly. 
This mode reduction has practical relevance for our statistical analysis
in Sec.~\ref{sub:powerspectrum}, where we correlate eigenmodes to compute the one-filament power spectrum --- an
application which we find to be intractable when $\mathcal{O}(10^4)$
eigen functions are involved.

Note that while mass conservation is not explicitly enforced in eq.~\eqref{eq:our_approach}, we find both coefficient estimators to satisfy $\sum_j |a_j|^2 = 0.99$
to within $1\%$ accuracy.\footnote{Recall that we only fit up to $R_{99}$.
}

\subsection{\label{sub:interference_in_filaments} Interference Properties in Reconstructed FDM Filaments}

Before we analyse statistical properties of a fuzzy filament population in Sec.~\ref{sec:statistics}, we close this section by highlighting some
qualitative and quantitative single filament properties in the presence of
interference.

\begin{figure}
    \centering
    \includegraphics[width=\columnwidth]{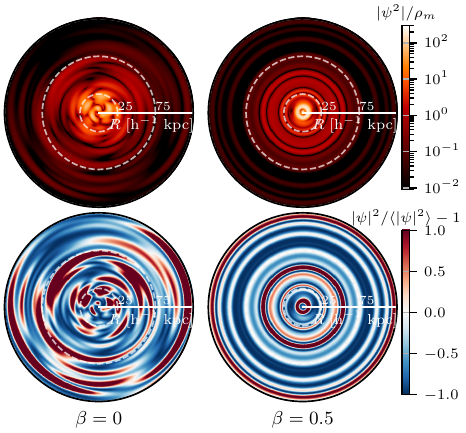}
    \caption{\label{fig:fringe_patterns}%
        (\textbf{Top row}) The density $|\psi|^2$, i.e. the static terms shown in Fig.~
        \ref{fig:density} \emph{and} the interference term, as a function of
        the anisotropy parameter. 
        (\textbf{Bottom row}) Interference modulates the density comparable
        to the magnitude of the background density leading to a concentric ring
        pattern for the radially biased case (right) and a more mixed
        interference fringe configuration under isotropic conditions (left). 
    }
\end{figure}

Figure~\ref{fig:fringe_patterns} illustrates the interference contained in the
optimised wave functions shown in Fig.~\ref{fig:density}. 
Strong oscillations around the background density are
found and the radially biased $\beta > 0$ velocity dispersion is reflected by a
concentric interference fringe pattern.

\begin{figure*}
    \centering
    \includegraphics[width=\textwidth]{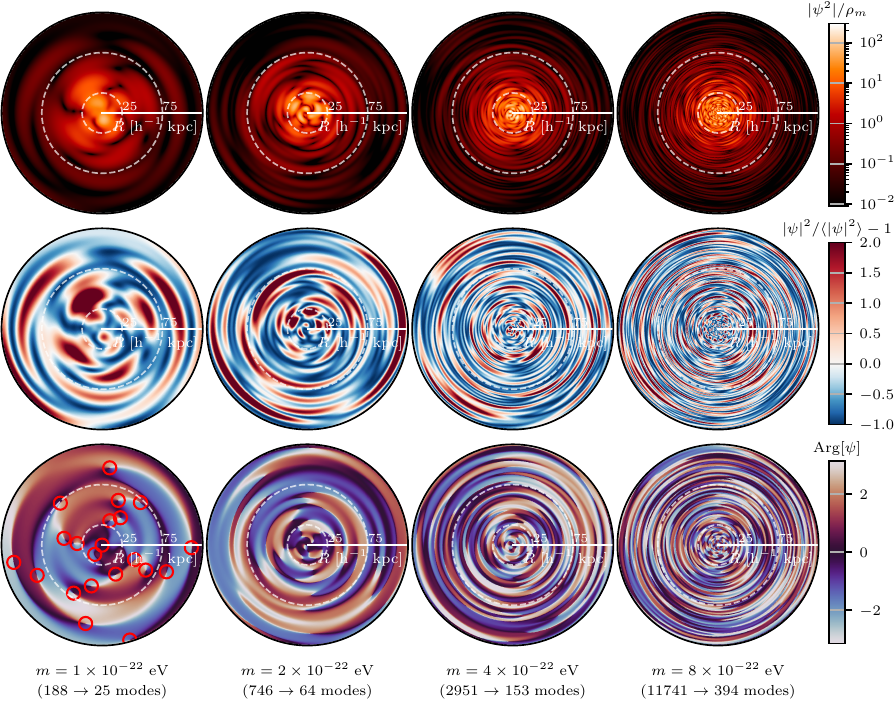}
    \caption{\label{fig:isotropic_crosssections}%
        Reconstructed FDM interference as a function of
        axion mass for an isotropic background density at redshift $z=4$.
        \textbf{(First Row)} 
        FDM density, $|\psi|^2$, as in eq.~\eqref{eq:interference}, i.e. including both the time static and 
        interference term.
        \textbf{(Second Row)} 
        The interference term relative to static background density.
        Strong, $\mathcal{O}(1)$, interference features are observable both in the radial and 
        azimuthal direction, indicative of an isotropic, $\beta = 0$, background
        density.
        \textbf{(Third Row)} 
        The wave function phase $\mathrm{Arg}[\psi]$. Quantized vortices (red
        circles, shown only for the left hand column), another feature unique for FDM, correspond to $2\pi$ discontinuities in the phase
        function, or as described in the main text, places of non-zero
        vorticity, eq.~\eqref{eq:vorticity}.
    }
\end{figure*}

In Fig.~\ref{fig:isotropic_crosssections} we depict a selection of filament
cross sections for an isotropic background density and a range of FDM masses
$\SI{1e-22}{\electronvolt}\leq m \leq \SI{8e-22}{\electronvolt}$ at $z=4$.

Evidently, as the particle mass increases the fringe spacing in both the radial and
azimuthal direction decreases, cf. first row. In all cases, however, the overall density is strongly
modulated and incurs $\mathcal{O}{(1)}$ changes with respect to the background, cf.
second row.

Since the wave function $\psi$ is the superposition of single-valued
eigenstates, $\psi$ is single-valued too. Consequently, the circulation $\Gamma$
related to the phase difference that arises from transporting the phase
$\Theta = \mathrm{Arg}[\psi]$ around any closed loop $\gamma$ must be quantized:
\begin{equation}
    \label{eq:circulation}
    \Gamma \propto \oint_\gamma \dd\Theta = 2\pi n\;\;\text{with}\;\; n \in \mathbb{Z} \;.
\end{equation}

If $\gamma$ encloses no roots of the wave function, the winding number $n$
vanishes and the dynamics within this region may be expressed in terms of
a potential flow $v \propto \nabla \Theta$ obeying the Madelung equations. 

In the presence of destructive interference, however, roots can occur and if $\gamma$ contains 
such loci --- so-called quantum vortices --- then $n>0$. 
Put differently, vortices act as discrete sources of
vorticity $\nabla \times v$ and interference can generate them. Since the
Madelung formulation makes $\nabla \times v = 0$ manifest, quantum vortices can
only arise if we solve eq.~\eqref{eq:nlse} or eq.~\eqref{eq:se} which pose no
restrictions on the value of the circulation except for eq.~\eqref{eq:circulation}. It is
thus no surprise that the wave functions of Fig.~\ref{fig:isotropic_crosssections} host interference-induced 
quantum vortices, which we locate for $m=\SI{1e-22}{\electronvolt}$ as
peaks in the vorticity,
\begin{equation}
    \label{eq:vorticity}
    \nabla \times v = (im)^{-1} \nabla \times \psi^*\nabla\psi\;,
\end{equation}
cf. red circles in the third row. As expected, these loci coincide with regions
where the phase $\Theta$ wraps around once, i.e. $n=1$.

An extensive account on theoretical and observational consequences of quantum
vortices for wave DM haloes is given in \cite{Hui2021}. For
filaments, \cite{Alexander2022} conjectured that a sufficiently large number
of vortex lines may generate a bulk rotation that is consistent with recent
observations \citep{Wang2021}. Reconstructing this vortex population according to the
prescription of the present work may be a way to test this hypothesis systematically.\footnote{%
    For a numerical analysis of vortices in haloes, see \cite{Liu2023}
}

\section{Statistical Aspects of FDM Filaments}
\label{sec:statistics}
With an idealised model for FDM filaments at hand, we turn our focus to
a statistical characterisation of fuzzy cosmic filaments. Section
\ref{sub:mass_function}
investigates general population statistics and derives the mass function for
FDM filaments. This will allow us to promote our
effectively two-dimensional, single cylinder considerations of Sec.~\ref{sub:interference_in_filaments} 
to a three dimensional filament population. 
Mathematical details for
the construction of the filament mass function  are deferred to Appendix~\ref{appendix:ellipsoidal_collapse}.
Section~\ref{sub:powerspectrum} analyses how the matter power spectrum is
modified in the presence of interference. Appendix~\ref{appendix:cylinder_power}
summarises its computation --- more precisely the one-filament contribution.

\subsection{\label{sub:mass_function} The Filament Mass Function}
The formation of the cosmic web constituents can be understood as a direct consequence of 
the ellipsoidal collapse model, e.g. \cite{White1979, Peebles1980, Bond1996}. 
Tidal forces shear spherical patches centred around the peaks of the
high-redshift Gaussian density field into a triaxial geometry. 
These initial, ellipsoidal overdensities continue to evolve under their 
self-gravity, external tides and Hubble expansion. 
Appendix~\ref{appendix:ellipsoidal_collapse} summarises relevant contributions in detail \citep{Bond1996}.

\begin{figure}
    \centering
    \includegraphics[width=\columnwidth]{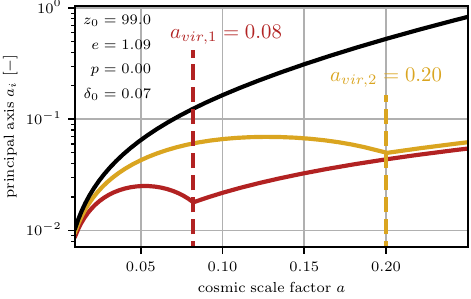}
    \caption{\label{fig:ellipsoidal_collapse}
        An example of ellipsoidal collapse in a $\Lambda$CDM background for a
        homogeneous ellipsoid with mass $M = \SI{5e9}{M_\odot}\,h^{-1}$. 
        The equation of motion, eq.~\eqref{eq:ellipsoidal_collapse},
        preserves the homogeneity of the initial conditions and it is thus
        sufficient to observe the evolution of the axis scale factors $a_i(a)$.
        After an initial expansion, each axis turns around, contracts and
        freezes out according to the steady-state tensor virial theorem, eq.~
        \eqref{eq:freeze_out}. The subsequent evolution is fixed to ensure a constant comoving
        axis size $R_i = a_i(a) R_\mathrm{ini}/a$. 
        We identify a filament as an object with two frozen axes ---
        in the depicted scenario realised at $z=4$. The residual triaxiality, i.e. the mismatch between the yellow and red lines at large \(a\), is
        ignored for the construction of our filament population.
    }
\end{figure}
We illustrate an example evolution in the principal axis system of the
ellipsoid for a $\Lambda$CDM background with concordance parameters
\citep{Planck2018} in Fig.~\ref{fig:ellipsoidal_collapse}. 
Depicted are the principal axes' scale factors
$a_i$ which relate to the physical axis size via the initial radius of the
undeformed, spherical Lagrangian space patch by $r(a)=a_i(a) R_\mathrm{ini}$.
The initial conditions were constructed for a filament of mass
$M=5\times10^9\;h^{-1}\text{ M}_\odot$ and tuned to allow for filament formation at 
$z=4$ \citep{Sheth2001} in the sense that we describe shortly. 
One may regard Fig.~\ref{fig:ellipsoidal_collapse} as
the idealised evolution of the smooth background density that leads to the
fuzzy filament shown in Fig.~\ref{fig:showcase_crosssection}.

The evolution proceeds in a self-similar, homogeneity-preserving fashion --- a
consequence of the quadratic nature of the ellipsoid's potential. 
Each axis undergoes a Hubble flow driven expansion phase, 
followed by a turn around, gravity-induced contraction and subsequent freeze-out.

Note that both the transition redshift from contraction to 
freeze-out and its subsequent evolution are not a direct consequence of the
equations of motion eq.~\eqref{eq:ellipsoidal_collapse}, but must be imposed by hand. A variety of approaches
exist depending on the sought-after properties of the post-freeze out state:
\cite{Bond1996, Sheth2001} halt and freeze the collapse of principal axis $i$
at $a_i(a)/a = f = 0.177$. 
This ensures that, after the last axis collapsed, the nonlinear density contrast is consistent 
with a virialized halo under spherical collapse. In an Einstein-de Sitter (EdS) background, this implies:
\begin{equation}
    \label{eq:adhoc_freeze_out}
    1+\delta(a) = \frac{a^3}{a_1 a_2 a_3} = f^{-3}= 178\;.
\end{equation}
\cite{Desjacques2008} analyses the dependence between ellipsoidal protohaloes 
and their environment and adopts the same freeze-out condition as 
\cite{Bond1996}, 
but makes angular momentum conservation subsequently manifest.
\cite{Angrick2010} note that the value of $f=0.177$ has no fundamental
motivation for non-spherical collapse and therefore apply the steady-state tensor viral theorem
to the evolving ellipsoidal density to derive a theoretically 
well-motivated per axis freeze-out radius, cf. eq.~\eqref{eq:freeze_out}.

In what follows, we adopt the approach that is most consistent with our
steady-state cylinder considerations of Sec.~\ref{sub:filament_cylinder}. We freeze each
axis once the tensor virial theorem is compatible with a steady-state
configuration \citep{Angrick2010}, cf. eq.~\eqref{eq:freeze_out}, and enforce a constant comoving radius afterwards.
\emph{
A filament is then identified as an object for which two out of the three 
principal axes are frozen} \citep{Shen2006, Fard2019} and we see that enforcing a constant comoving axis-length
reduces the triaxiality at filament formation so that a cylindrical
approximation becomes viable.

How many of such defined filaments do we expect per comoving
volume at formation redshift $z$? The excursion set formalism 
\citep{Bond1991} provides an answer to this given two additional ingredients.
The filament mass variance $\sigma^2(M)$, eq.~\eqref{eq:variance}, and
$\delta_\mathrm{ec}(\sigma^2, z)$, i.e. the linearly extrapolated density contrast at the time of filament formation according to ellipsoidal collapse evolution.
Here we only give a brief account of the choices made to compute both
quantities. We refer to Appendix~\ref{appendix:ellipsoidal_collapse} for
mathematical details.

The mass variance $\sigma^2(M)$ results from smoothing the cosmic density field via a window
function of spatial scale $R$ (or equivalently mass scale $M$), cf. eq.~\eqref{eq:variance}. 
The suppression properties of FDM are encoded in
$\sigma^2(M)$ by virtue of its linear power spectrum entering the
smoothing operation. The translation from spatial scale $R$ to filament
mass $M$ depends on the choice of window function, which must be sufficiently sharp
in $k$-space to ensure that the mass function remains sensitive to suppressive features encoded in the 
matter power spectrum \citep{Schneider2013}.
In accordance with
\cite{Du2023}, we adopt a calibrated smooth-$k$ filter \citep{Leo2018,
Bohr2021} for this task. In the end, we restrict our analysis to filament masses unaffected by the choice of filter.

For the linear density contrast at filament formation
$\delta_\mathrm{ec}(\sigma^2, z)$, one commonly employs fitting formulae derived from ellipsoidal 
collapse studies. \cite{Shen2006}
provide such a fit in the case of an EdS background and ad-hoc freeze-out condition
eq.~\eqref{eq:adhoc_freeze_out}. Since we are not aware of a similar result
for the virial freeze-out condition eq.~\eqref{eq:freeze_out} in $\Lambda$CDM, 
we determine $\delta_\mathrm{ec}(\sigma^2, z)$ by optimising
the initial ellipsoidal over density $\delta(a_\text{ini}\mid M)$, until we are
able to form filaments of mass $M$ at redshift $z$. Linear theory is then used to extrapolate
forward again, i.e.:
\begin{equation}
    \delta_\mathrm{ec}\left(\sigma^2, z\right) =
    \frac{D(z)}{D(z_\mathrm{ini})}\delta\left(a_\text{ini}\mid M(\sigma^2)\right)
    \;,\quad z_\mathrm{ini} = 100\;,
\end{equation}
with $D(z)$ as the $\Lambda$CDM growth factor \citep{Dodelson2003}.
Scale-dependent FDM growth is irrelevant on the filament scales that we consider here.

The number of filaments with masses between $M$ and $M+\text{d}M$ at redshift 
$z$, i.e., the filament mass function (FMF), then follows from:
\begin{equation}
    \label{eq:fmf_def}
    n(M)\,\text{d}M = \frac{\rho_{m0}}{M}f_{B}(\sigma^2)\,\text{d}\sigma^2\,,
\end{equation}
with $\rho_{m0}$ and $f_{B}$ as present-day matter density and multiplicity 
function of the barrier,
\begin{equation}
    B(\sigma^2, z) = \frac{D(z)}{D(0)}\delta_\mathrm{ec}(\sigma^2,z)\;,
\end{equation}
respectively. Details on the computation of the multiplicity function are deferred to Appendix~\ref{appendix:ellipsoidal_collapse}.

\begin{figure}
    \centering
    \includegraphics[width=\columnwidth]{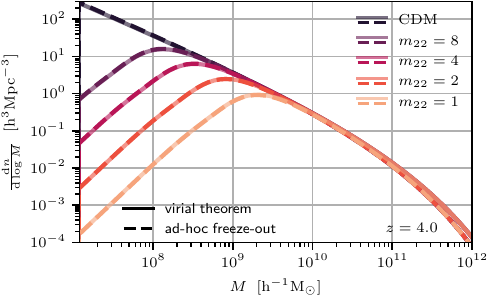}
    \caption{\label{fig:filament_mass_function}
        Filament mass function (FMF) for various FDM masses at redshift $z=4$.
        The suppression in the linear FDM power spectrum translates to a power
        suppression for filament masses. For filament masses
        $M>3\times10^9 h^{-1}\text{ M}_\odot$, all considered FDM cases are
        indistinguishable from CDM and it is in this regime in which we sample
        the FMF in order to guarantee the same overall number density and thus
        same significance of interference effects measured in the correlation
        functions discussed in Sec.~\ref{sec:statistics}.
    }
\end{figure}

Figure~\ref{fig:filament_mass_function} depicts the FMF
for a selection of FDM masses at redshift $z=4$. The low-mass suppression is the
result of the small-scale suppression of matter power due to the presence of quantum
pressure --- a property of the diffusive, kinetic term in \eqref{eq:nlse}. 
We also show the FMF obtained if the freeze-out condition
eq.~\eqref{eq:adhoc_freeze_out} (dashed) is used instead of eq.~\eqref{eq:freeze_out} (solid). The dashed FMF can be understood as the $\Lambda$CDM
variant of the FMF derived in \cite{Shen2006}. We find no deviations
for $M<10^{11}\;h^{-1} \text{ M}_\odot$.

The FMF together with the geometry of the final ellipsoidal state may
also be used to promote our effectively two-dimensional, single filament
considerations of Sec.~\ref{sec:method} to a simplified, three-dimensional filament 
population,
To this end, we translate the total filament mass into initial conditions
for the ellipsoidal collapse \citep{Sheth2001}, follow the principal axis
evolution until $z=4$ and map the strongly prolate, final ellipsoid to a cylindrical geometry. 
More precisely, we set:
\begin{align}
    L(M)&\equiv 2a_3(z)R_\mathrm{ini}(M)(1+z)^{-1}\;,\\
    R_{90}(M)&\equiv\sqrt{a_1(z)a_2(R)}R_\mathrm{ini}(M)(1+z)^{-1}\,,
\end{align}
as the comoving length and radius containing $90\%$ of the total line mass
$\mu$. 
Equation~\eqref{eq:line_mass_dispersion} then provides the 
per-cylinder velocity dispersion $\sigma^2$ if we identify $\mu \equiv M/L$.
With these parameters at hand, we construct the eigenstate library, compute
the phase space distribution and optimise the wave function coefficients to
arrive at a semi-analytical expression for $\psi$. 

Next, we assume a fiducial, longitudinal profile of the form:
\begin{equation}
    \label{eq:lambda}
    \lambda(Z\mid L) = \frac{1}{2}\left[
        \mathrm{erf}\left(\Delta\left(z + \frac{L}{2}\right)\right)
        -
        \mathrm{erf}\left(\Delta\left(z - \frac{L}{2}\right)\right)
    \right] \;,\;
    \Delta = \frac{10}{L}\,,
\end{equation}
such that
\begin{equation}
    \label{eq:u_FDM}
    \begin{split}
        \rho_\text{CDM}\left(\bm x \mid M\right) &= \lambda\left(Z \mid L(M)\right)\rho_\text{BG}(R)\;,\\
        \rho_\text{FDM}\left(\bm x \mid M\right) &= \lambda\left(Z \mid L(M)\right)|\psi(R, \Phi)|^2
    \end{split}
\end{equation}
constitute the CDM/FDM filament densities, smoothly restricted to
$Z=\pm L/2$ along the filament spine.

\begin{figure*}
    \centering
    \includegraphics[width=\textwidth]{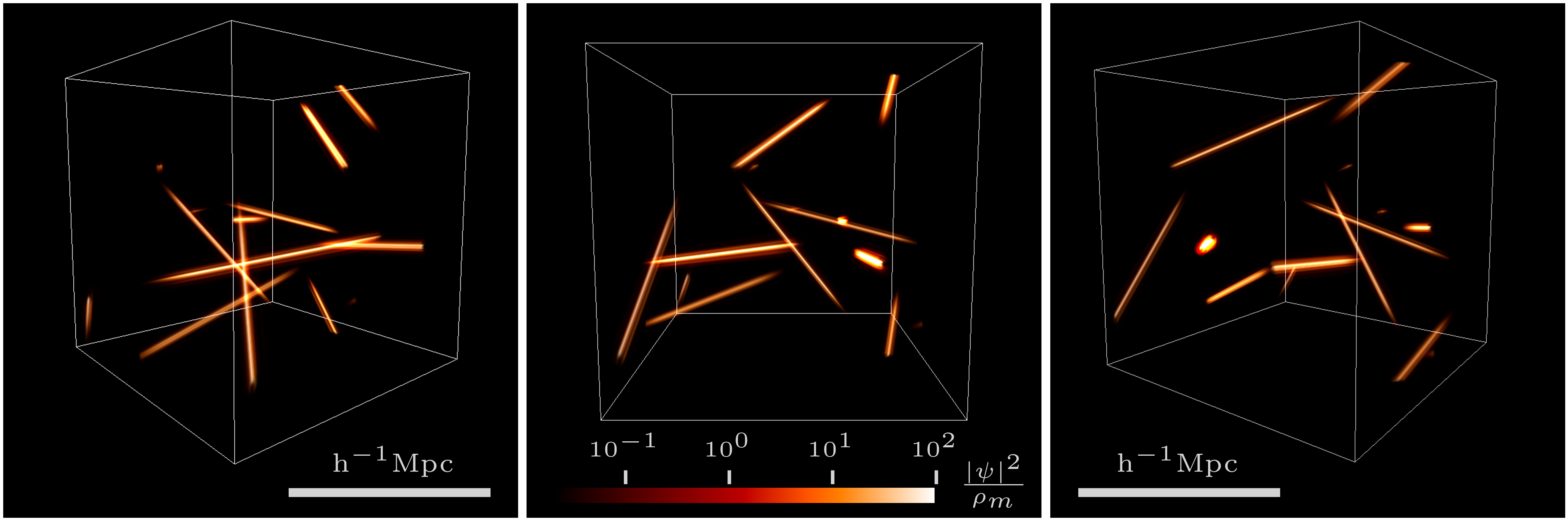}
    \caption{\label{fig:cylinder_gas_in_box}%
        Volume rendering of the idealised FDM filament sample for
        $m_{22}=1$ in a periodic, comoving box of
        $L=2\;h^{-1} \text{ Mpc}$ at $z=4$, generated by
        sampling the FMF in Fig.~\ref{fig:filament_mass_function} 
        for masses $M>3\times10^9\;h^{-1} \text{ M}_\odot$. 
        The characteristic filament length is $\simeq 1-2\;h^{-1} \text{ Mpc}$.
        The random iid placement of the cylinders neglects any
        filament-filament cross-correlations encoded in the cosmic web.
        Consequently, one expects the power spectrum of the box shown to be identical to the 
        one-filament term in eq.
        \eqref{eq:P1f}. This is supported by Fig.~\ref{fig:powerspectra} which
        compares the spectrum of the domain depicted in this
        Figure (dashed, yellow) with the stacked spectra following from eq.
        \eqref{eq:P1f} (solid, blue).
    }
\end{figure*}

Figure~\ref{fig:cylinder_gas_in_box} illustrates an example realisation of our
filament population assuming cylinder locations 
$\{\bm x_i\}_{i=1\dots N}$ 
and orientations $\bm \{\bm d_i\}_{i=1\dots N}$ follow from:
\begin{equation}
    \bm x_i \sim U([0, L]^3)\;,\; \bm d_i \sim S^2 \;.
\end{equation}
Samples are accepted if no cylinders overlap in the fundamental
cell and all its periodic extensions, which would violate
our assumption of all filaments being gravitationally isolated.
This clearly ignores any large scale filament-filament cross-correlation
imprinted on the cosmic web (which, in analogy to the halo model, at leading order should be given by the linear power spectrum).
More refined placement techniques, in particular a peak-patch description \citep{Bond1996}
may be used to alleviate this shortcoming. 

\subsection{\label{sub:powerspectrum}The Matter Power Spectrum}
How are two-point density correlations impacted by the presence of interference
in filaments? To answer this, we turn to the isotropic power spectrum:
\begin{equation}
    \label{eq:powerspectrum}
    P(k) \equiv \left\langle |\hat \delta(\bm k)|^2 \right\rangle_{|\bm k|=k}
    \;.
\end{equation}
Fully-fledged cosmological simulations of FDM report an excess
correlation in the FDM matter power spectrum compared to its CDM counterpart for highly 
non-linear $k>\mathcal{O}(100)\;h\text{ Mpc}^{-1}$
\citep{Veltmaat2016,Mocz2020, May2021, May2022,Lague2023}, and conjectured, quite intuitively, this
power boost to originate from interference fringes imprinted on the density
field $|\psi|^2$.
The wave function reconstruction scheme of Sec.~\ref{sec:method} combined with the population statistics
developed in Sec.~\ref{sub:mass_function} allow us to test if quasi-virial filaments can in principle be the
source of such a power boost.

Interference is a local phenomenon, i.e. the result
of trapped wave modes beating against each other in a common gravitational
potential sourced by a single filament. One should not expect
any correlation from the interference dynamics between sufficiently separated
filaments. Sec.~\ref{sec:method} makes this manifest by treating the wave function reconstruction ex situ.

Motivated by the standard halo model \citep{Cooray2002}, we therefore
decompose eq.~\eqref{eq:powerspectrum} into a
one-filament, $P^{\mathrm{1f}}(k)$, and two-filament contribution,
$P^{\mathrm{2f}}(k)$. The two-filament term captures correlation between filaments. 
We ignore this in the present model, since the cross-correlation of filaments will not 
contain an interference term, and as already noted at leading order is captured by the 
linear power spectrum.

The one filament term reads:
\begin{equation}
    \label{eq:P1f}
    P^{1f}(k) =
    \frac{1}{\bar\rho_m^2}\int\text{d}M\, M^2n(m)P_{\mathrm{single}}(k\mid M),
    \;\; \bar\rho_m = \int \text{d}M M n(M) \;,
\end{equation}
and measures the spatial correlation at the ensemble level by stacking
isotropised spectra of individual filaments:
\begin{align}
    \label{eq:Psingle}
    P_{\mathrm{single}}(k\mid M) &= \left\langle |\hat u(\bm k \mid
    M)|^2\right\rangle_{|\bm k| = k}\;,\\
    \hat u &= \frac{1}{M}\int\text{d}^3\bm x\;e^{i\bm k \cdot \bm x}\rho\left(\bm x \mid M\right)\;,
\end{align}
of total mass $M$.

\begin{figure}
    \centering
    \includegraphics[width=\columnwidth]{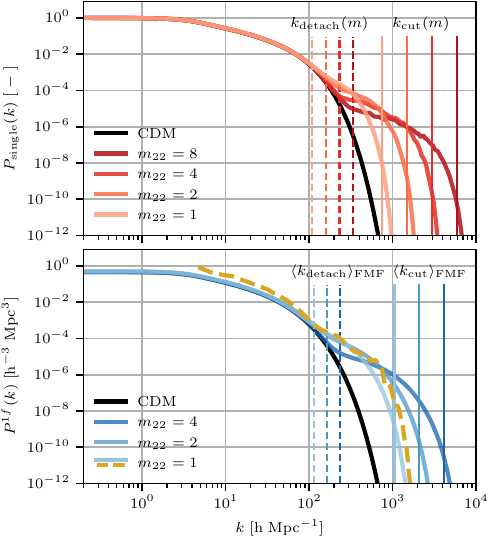}
    \caption{\label{fig:powerspectra}%
        Three-dimensional filament matter power spectra of the interference-free CDM background and 
        the reconstructed FDM density for a selection of FDM masses at $z=4$. 
        \textbf{(Top)}
        Single filament spectrum, $P_\text{single}(k\mid M)$, for an example
        filament of total mass $M~=~5\times10^9\;h^{-1} \text{ M}_\odot$.
        One finds all FDM spectra to be compatible with CDM
        down to the spatial extent of the ground state, $R_\text{gs}\equiv\bra{\psi_0}R\ket{\psi_{0}}$ (vertical, dashed line) --- effectively the
        characteristic scale of the cylinder. 
        Once sufficiently small scales inside the filament are probed, the interference
        contribution to $|\psi|^2$ enhances the two point density correlation
        and boosts $P_\text{single}(k\mid M)$ above the CDM baseline. Note that
        for larger $m_{22}$: (i) $P_\text{single}(k\mid M)$ detaches later from
        the CDM baseline, (ii) the boost is shallower but (iii) extends to higher $k$ until the 
        suppression scale $k_\text{cut} \propto m$ is reached (vertical, solid lines).
        \textbf{(Bottom)}
        One-filament term $P^{1f}(k)$ --- a stacked version of many
        $P_\text{single}(k\mid M)$, weighted according to the FMF in Fig.~\ref{fig:filament_mass_function}. 
        We limit the filament population to
        the mass window $M\in\left[3\times 10^9,
        3\times10^{10}\right]\;h^{-1}\text{ M}_\odot$ and compute the one
        filament term either directly from the semi-analytical expression of the
        wave function developed in Appendix~\ref{appendix:cylinder_power} (solid,
        blue)
        or by populating a box as shown in Fig.~\ref{fig:cylinder_gas_in_box}
        (dashed, yellow). All quantitative observations made for
        the single cylinder case translate to $P^{1f}(k)$ after a FMF-weighted
        average according to eq.~\eqref{eq:fmf_q}, i.e. $R_\text{gs} \to \langle
        R_\text{gs}\rangle_\text{FMF}$ (vertical, dashed lines) and $k_\text{cut} \to \langle
        k_\text{cut}\rangle_\text{FMF}$ (vertical, solid lines).
    }
\end{figure}

For the density depicted in Fig.~\ref{fig:cylinder_gas_in_box}, we expect
$P(k) = P^{1f}(k)$ since filament-filament cross-correlations are suppressed by
construction. This is supported by computing $P(k)$ for the box shown in Fig.~\ref{fig:cylinder_gas_in_box} (dashed) and comparing it against $P^{1f}(k)$ from eq.
\eqref{eq:P1f} (solid) in the bottom panel of Fig.~\ref{fig:powerspectra}. 
Thus, populating a three dimensional box as described in Sec.~\ref{sub:mass_function} and subsequent $k$-space binning is a simple way to compute 
the one-filament term. However, the obvious tension between a large domain (better
statistics due to more filaments) and sufficient resolution to resolve the
finest interference fringes gets quickly intractable as the FDM mass increases.

Fortunately, we have access to $\psi$ and have explicitly reduced its library
size. Therefore, it is advantageous to compute $P^{1f}$ according to eq.~\eqref{eq:P1f}, i.e., by directly stacking a large number of single-filament spectra, 
$P_{\mathrm{single}}(k|M)$. This approach also allows us to consider the effects of 
interference at the level of individual filaments before generalizing to $P^{1f}(k)$. 
Technical details are provided in Appendix~\ref{appendix:cylinder_power}.

The top panel of Fig.~\ref{fig:powerspectra} depicts the single filament power spectrum
$P_{\mathrm{single}}\left(k \mid M=5\times10^9\;h^{-1} \text{ M}_\odot\right)$,
for a selection of FDM masses (shades of red) alongside the interference free CDM case (black).

Comparing the FDM spectra with the CDM spectrum, we find a boost in matter
power for $k>100\;h \text{ kpc}^{-1}$, which we can attribute exclusively to
the presence of the interference cross-terms in $|\psi|^2$. 

In all considered cases, the FDM spectrum detaches
from the CDM baseline at some wavenumber $k_\text{detach}$, once the reciprocal correlation wavenumber $k$ starts to probe scales
interior to the filament. The exact value of $k_\text{detach}$ is observed to
have a weak, monotonically increasing, dependence in $m_{22}$ and we find it to be reasonably well
approximated by the expectation value of the radial position operator with
respect to the ground state mode,  i.e. 
$k_\text{detach}\propto\bra{\psi_0}R\ket{\psi_{0}}^{-1}$ (dashed, vertical lines).
This, together with the shallower boost for higher FDM mass, is consistent with our 
expectation of recovering the CDM scenario in the limit $m_{22}\to\infty$.

The power boost extends to higher $k$ until a FDM mass-dependent cut-off scale
$k_\text{cut}(m)$ drives
$P_\text{single}(k)$ to zero. We interpret this scale as a non-linear extension
of the linear Jeans suppression scale set by the uncertainty principle
$u_\text{cut} x_\mathrm{cut} \gtrsim m^{-1} \hbar$. Figure~\ref{fig:powerspectra} suggests $k_\text{cut}(m) \propto m$ such
that the conjugate velocity $u_\mathrm{cut}$ entering the uncertainty relation
must be mass independent. As most interference is located in the filament
outskirts where high energy/angular momenta modes dominate, it is intuitive to
associate $u_\text{cut}$ with the FDM mass independent, WKB behaviour of the mode coefficients
$|a_{nl}|^2 \propto f(E)$. We therefore return to the isotropic phase space distribution shown in Fig.~\ref{fig:df} (black, dashed line), which for $E/V(R_{99}) > 0.1$ is well described by an
isothermal distribution \citep{Binney2008} of the form:
\begin{equation}
    f(E) \propto \exp{\left(-\frac{E}{\sigma_u^2}\right)} \quad\Rightarrow\quad
    \langle u^2\rangle = \frac{\int\text{d}u\,u^3 f\left(E(u)\right)}{\int \text{d}u\,u
    f\left(E(u)\right)} = 2 \sigma_u^2\,.
\end{equation}
Fitting the isothermal ansatz to Fig.~\ref{fig:df} and setting $u_\text{cut} =
\sqrt{\langle u^2 \rangle}$ results in the solid, vertical lines shown in Fig.~\ref{fig:powerspectra} which are in satisfying agreement with the suppression cut-off of the 
interference boost.

Figure~\ref{fig:powerspectra}'s bottom panel depicts the one-filament term $P^{1f}(k)$ for 
$m_{22}=1,2,4$ restricted to an ensemble with mass $M\in\left[3\times 10^9, 3\times10^{10}\right]
\;h^{-1} \text{ M}_\odot$. The lower total mass limit is chosen such that CDM and
all considered FDM cases have identical population statistics according to the
suppression physics encoded in the linear power spectrum, cf. Fig.~\ref{fig:filament_mass_function}. The upper limit is set by resource
constraints.\footnote{
    Recall that $P_\text{single}$ now needs to be computed for a range of total
    masses $M$. Higher mass filaments translate to deeper potential wells,
    imply more extensive eigenstate libraries and thus require us to solve more challenging 
    optimisation problems. Although parallelisation strategies exist,
    we have not put significant effort into making the latter scalable.
}

Let $\langle Q\rangle_{\text{FMF}}$ denote the FMF-weighted average of
quantity $Q$:
\begin{equation}
    \label{eq:fmf_q}
    \langle Q\rangle_{\text{FMF}} 
    \equiv 
    \int_{M_{min}}^{M_{max}} \text{d}M\; M^2 n(M) Q(M) \;.
\end{equation}
We then recover the same qualitative behaviour as seen in
$P_\text{single}(k)$ upon applying an FMF-average, i.e. one finds large scale equivalence with CDM, 
departure once scales smaller than $\left\langle
\bra{\psi_0}R\ket{\psi_0}\right\rangle_{\text{FMF}}$ are reached, and a sustained boost 
in power up to $\left\langle k_\mathrm{cut}(m)\right\rangle_{\text{FMF}}$.

\section{Discussion}
\label{sec:discussion}
Let us put these results into context.
As stated in our introductory remarks, interference, not just in filaments but in all 
constituents of the cosmic web, has been 
identified as a smoking gun signature since the advent of FDM simulations \citep{Schive2014}. 
State-of-the-art SP numerics allows us to follow FDM structure formation
from the kpc scale up to boxes of size $L=10\;h^{-1} \text{ Mpc}$ and
down to $z\gtrsim 0$. This is sufficient to resolve the impact of wave phenomena
on summary statistics for a canonical FDM mass of 
$m_{22} < 1$ \citep{May2021, May2022}. For the matter power spectrum,
the wave-like signatures may be summarised as follows.

For sufficiently high redshift, $z\geq5$, the difference
between CDM and FDM dynamics is negligible compared to the influence of the initial
conditions \citep{May2021, May2022} --- the classical FDM approximation is thus
justified in this regime. At lower redshift, however, the impact of FDM's wave-like evolution 
is more significant and introduces two additional features in $P(k)$ compared to
its CDM counterpart and irrespective of the choice of initial conditions: 
(i) larger scales, $k\lesssim \mathcal{O}(10)\;h\text{ Mpc}^{-1}$, are suppressed and 
(ii) smaller scales, $k\gtrsim \mathcal{O}(100)\;h\text{ Mpc}^{-1}$, experience an enhancement.

Our proof-of-concept analysis indicates that interference induced by a steady-state fuzzy 
filament population can produce the observed power boost on scales broadly consistent with
\cite{Mocz2019, May2021, May2022}. The impact of wave dynamics on larger,
quasi-linear scales is not accessible by our model. Intuitively, one would expect
their evolution to be better described perturbatively, i.e. by means of a wave
equivalent of Zeldovich's approximation, that captures the coherent flow of matter
waves, rather than the quasi-virial limit our filament proxies assume.

\cite{Uhlemann2019} provide such a description via the propagator perturbation 
technique (PPT). From the
perspective of interference/wave modelling, the crucial
difference between PPT and classical Lagrangian methods (LPT) is the
representation of the cosmic density as a semiclassical wave function
rather than a set of fluid test particles.\footnote{
    In fact, to leading order, PPT is equivalent to free wave propagation.
}
Thus, wave effects are retained in PPT up to mildly nonlinear scales.

\cite{Gough2024} deployed PPT to
analyse the impact of wave effects, i.e. interference and quantum pressure, 
on summary statistics for spatial scales accessible to a perturbative treatment. 
In case of the matter power spectrum at
$z=4$, wave-like PPT evolution produced a suppression of power for scales 
$k\lesssim 10\;h \text{ Mpc}^{-1}$ compared to the CDM LPT dynamics.
This result was found to be in broad agreement with the $m_{22}=0.1$ SP
simulations of \cite{May2022}.

Together, PPT and the interference reconstruction of this work thus provide a
complementary picture that is in qualitative accordance with simulation results
and suggests the following intuitive picture.
Up to mildly-nonlinear scales, FDM flows coherently into the gravitational wells
of over dense regions and quantum pressure suppresses structure, even if
interference is taken into account. Non-linear scales, by contrast, are boosted by
the interference emerging in objects such as filaments until Heisenberg's uncertainty suppresses 
structure growth entirely.

At present, observed quasar absorption spectra allow us to probe $P(k)$ down to 
$k\simeq \mathcal{O}(10)\;h \text{ Mpc}^{-1}$ \citep{Boera2018}, while interference
from our filament population enhances scales at least two orders of magnitude
smaller and way past the filtering scale $k_f\simeq 40\ h \text{ Mpc}^{-1}$
\citep{Hui1997}. We
conclude that steady-state filaments, as described in this work, are
not expected to affect ultralight DM constraints from the Ly$\alpha$ forest 
\citep[e.g.][]{Irsic2017, Kobayashi2017, Rogers2021} 
due to a degeneracy between linear quantum pressure suppression and interference enhancement.
This further justifies the use of the "classical FDM" approximation, i.e. FDM
initial conditions for CDM simulations, for modelling mildly-nonlinear scales
probed by the Ly$\alpha$ forest. 

Interestingly, the mass dependence of $k_\text{detach}$ and $k_\text{cut}$ 
indicates that axions lighter than those considered in this work may be sensitive to this degeneracy
at larger spatial scales. At the same time, a lower value of $m_{22}$ implies fewer eigenstates and 
consequently reduced interference enhancement at fixed $k$ relative to the CDM baseline. 
Focusing on higher-mass filaments at redshifts $z<4$ may help maintain the amplitude of the 
interference bump, while shifting to scales accessible to large-scale structure probes.

A natural question is to what extent interference in virialised, spherically
symmetric haloes contributes to small scale modifications in $P(k)$. Answering
this in detail is beyond the scope of this work. 
Here, we only state that at high redshift, $z>3$, haloes were found
to contribute more than an order of magnitude less bound mass to the 
cosmic web than filaments \citep{Dome2023b}. This should translate to a
suppressed halo mass function relative to the FMF and consequently
a reduced significance of $P^{1h}$ compared to $P^{1f}$, cf. eq. \eqref{eq:P1f}.
At the level of individual filaments, the matching symmetry between the
\emph{isotropised} power spectrum and the 
\emph{spherically symmetric} wave function results in cross-term
cancellation for certain angular momentum combinations.\footnote{%
    More precisely, we find $\langle |u(\bm k \mid M)|^2\rangle_{|\bm k| = k}$
    for a halo to depend on the Wigner-3j symbols for which selection rules
    apply.
}
Moreover, halos are more compact and occupy less volume than filaments
\citep{Dome2023b}. We therefore expect a higher value of $k_\mathrm{detach}$. Higher
characteristic velocities inside halos translate to higher values of 
$k_\mathrm{cut}$. In conclusion, we anticipate a power boost due to halo interference
to be present at smaller scales and significantly reduced amplitude compared to
filaments.

\section{Conclusion \& Prospects}
\label{sec:conclusion}

In this work, we took initial steps towards a more principled understanding of
the modelling and measurability of interference features in FDM filaments. To
this end:
\begin{itemize}
    \item We approximated FDM filaments as infinitely long, cylindrical objects,
        computed their associated eigenstates and superposed these states such that
        a (non)-isotropic, isothermal, steady-state solution of the classical Jeans equation is
        recovered. To reduce the complexity of the reconstructed wave function, 
        we computed a self-consistent phase space model for cold filaments and
        leveraged the WKB correspondence between mode coefficients and
        phase-space distribution to perform a data-driven mode selection. 
        We find the interference sourced by our simplified filament model to be
        in qualitative accordance with filament interference found in full SP
        simulations. 
    \item We applied the ellipsoidal collapse model alongside a virial freeze-out
        condition to construct a fuzzy filament mass function. 
        Combined with the geometry of the partially collapsed ellipsoid, 
        this approach allowed us to build a toy filament population that neglects 
        filament-filament cross-correlations, but incorporates interference fringes 
        generated by eigenstate cross-terms.
    \item Using this toy population, we computed the \emph{one-filament matter power spectrum} 
        for various FDM masses. Our proof-of-concept analysis revealed a boost in power 
        between the spatial scale associated with the ground state mode and the scale 
        set by Heisenberg's uncertainty principle at the root mean square velocity 
        of the classical phase-space distribution. 
        These findings are consistent with fully-fledged SP simulations
        \citep{Mocz2019, May2021, May2022} and complement 
        interference-sensitive, perturbative treatments of FDM structure
        formation \citep{Gough2024}.
\end{itemize}

Our work opens several avenues for extension. Below, we outline an incomplete list 
of directions:

\begin{itemize}
\item Borne out of theoretical simplicity
    \citep{Stodolkiewicz1963,Ostriker1964,Eisenstein1997} and numerical evidence
    \citep{Ramsoey2021}, we focused on an isothermal, steady-state approximation for filaments. 
    However, a deeper understanding of the dynamical state of FDM filaments remains elusive.
    An ideal test bed to make progress on this question is the
    fuzzy filament catalogue of \cite{May2022}. Assessing to which degree
    steady-state conditions are attained on a per filament basis may give insights into the
    relative abundance of this subpopulation and is thus a natural next step.
    Moreover, stacking the catalogue would provide access to the one-filament spectrum which
    could be directly compared with our bottom-up model.

\item An intriguing prospect is to merge PPT and interference reconstruction. 
    This may be achieved by combining PPT with the aforementioned
    peak-patch algorithm \citep{Bond1996} to propagate FDM initial conditions to a target redshift,
    smooth the resulting density field on mass scale $M$ and identify peaks above the 
    critical density suggested by ellipsoidal collapse. 
    Upon replacing these patches with the fuzzy cylinders developed in this
    work, one arrives at a density field with consistent large scale
    suppression, small scale enhancement of structure and correct
    filament-filament cross-correlation.

\item We have alluded to the possibility of using the presented model as
    a $N$-body simulation post-processing tool, in which particle positions and
    velocities inside filaments may be used to reconstruct the wave function (see
    the discussion around eq.~\eqref{eq:poisson_process}) and phase space
    distribution \citep{Schwarzschild1979}.

\item Beyond the semi-analytical treatment in this work, a
    Gaussian beam decomposition of $\psi$ and subsequent
    integration of the beam trajectories \citep{Schwabe2022} is appealing for
    non-stationary conditions. In
    this approach, phase information is retained and carried along 
    the travelling beams such that interference can be reconstructed locally by
    summing over all beams.
\item Having identified $P^{1f}$ --- the stacked single filament power spectrum
    --- as a statistic sensitive to interference fringes, we speculate that the
    weak lensing signal of stacked filaments may be used to reconstruct
    the surface mass density \citep[e.g.][]{Lokken2024} and thus the one-filament power spectrum
    observationally. This may pave the way for a novel ultralight DM mass limit
    based on interference alone.
    \footnote{%
        It is interesting to note in this context that strong lensing can probe 
        the existence of an interference substructure in radio arcs, 
        providing an effective way to derive mass limits \citep{Powell2023}.
    }
\item \cite{Gough2024} advocate to study the implications of interference on
    statistics beyond two-point functions. 
    A particularly interesting observable is the line 
    correlation function (LCF) \citep{Obreschkow2013, Wolstenhulme2015, Eggemeier2015}.
    The LCF is a regularised version of the three-point correlator 
    $\left\langle\hat \epsilon(\bm t -\bm x)\hat \epsilon(\bm r)\hat \epsilon(\bm t + \bm x)
    \right\rangle_{\bm t, |\bm x| = R}$,
    designed to capture information contained in the
    phases, $\hat \epsilon(\bm k) =\hat \delta(\bm k)/|\hat \delta(\bm k)|$, of the 
    density field.
    Not only does the LCF quantify statistical information
    absent in the power spectrum, it also measures the degree of straight filamentarity on the
    spatial scale $R$ \citep{Obreschkow2013}. Therefore, we anticipate the LCF to
    be an intriguing summary statistic able to quantify interference in filaments. 
    We have already done some preliminary work on this subject and refer to the
    \texttt{github} repositories below for an unoptimised Python implementation.
\end{itemize}
    
\noindent \textbf{\texttt{GitHub}}: The code used to generate the results in this
work can be found at
\href{https://github.com/timzimm/fuzzylli}{\faGithub\;timzimm/fuzzylli} and 
\href{https://github.com/timzimm/fdm_filaments}{\faGithub\;timzimm/fdm\_filaments}.

\begin{acknowledgements}
We thank Alexander Eggemeier for useful discussions on the (anisotropic) line
correlation function, Charis Pooni for
valuable comments on the wave function reconstruction, as well as Simon May,
Tibor Dome, Jens Niemeyer, Ren\'{e}e Hlo\v{z}ek and Philip Mocz for enlightening comments on interference in FDM filaments. 
TZ would like to thank the Theoretical Particle Physics \& Cosmology (TPPC) Group at
King's College London for their hospitality during the research stay, during
which parts of this work were carried out. TZ acknowledges financial support
through a Kristine Bonnevie stipend (University of Oslo). 
DJEM is supported by an Ernest Rutherford Fellowship from the Science and Technologies Facilities Council (ST/T004037/1). KKR is supported by an Ernest Rutherford Fellowship from the Science and Technologies Facilities Council (ST/Z510191/1). The Dunlap Institute is funded through an endowment established by the David Dunlap family and the University of Toronto.
This project has received funding from the European Union's Horizon 2020 research and innovation 
programme under the Marie Sk\l{}odowska-Curie grant agreement No. 945371.
\end{acknowledgements}

\bibliographystyle{aa}
\bibliography{filaments}

\begin{appendix} 

\section{Numerical Aspects of the Eigenstate Library Construction}
\label{appendix:numerics} 
The numerical construction of the radial eigenmodes for cylindrical
filaments is slightly more involved than the equivalent problem for haloes
and spherical symmetry. Here, we elaborate on important numerical aspects and
highlight differences to the halo case.

Recall the continuous eigen problem discussed in the main text:
\begin{equation}
    \label{eq:eigenproblem}
\begin{split}
    E_{nl} u_{nl}
    &= -\epsilon^2\left( \partial^2_R + \frac{1/4 - l^2}{R^2}\right)
    u_{nl} + V(R) u_{nl}\quad R\in\mathbb{R}^+\;,\\
    u_{nl}(0) &= u_{nl}(\infty) = 0\;,
\end{split} 
\end{equation}
$\epsilon^2 \equiv \hbar^2 /(2m^2a^2)$ and $u_{nl} = \sqrt{R} R_{nl}$ 
--- a transformation used to remove the first
derivative in the Laplacian and to make the angular momentum barrier appear
explicitly in the Hamiltonian. 

Since mode $u_{nl}$ decays exponentially in the
classically forbidden region:
\begin{equation}
    \label{eq:classically_forbidden}
\mathcal{F}=
\left\{R \in \mathbb{R}^+ \mid
    V_{\mathrm{eff},\psi}(R)~\equiv~V(R)+\epsilon^2\frac{l^2-1/4}{R^{2}} >
E\right\}\;,
\end{equation}
it suffices to seek for a solution to eq.~\eqref{eq:eigenproblem} on a truncated interval $R\in[0,
R_\mathrm{max}]$, where the endpoint $R_\mathrm{max}$ is fixed according to the asymptotic
behaviour of $u_{nl}$. As 
\begin{equation*}
    u_{nl}(R) \sim
    \exp\left(-\frac{1}{\epsilon}\int_{\max\mathcal{F}}^R\dd s\sqrt{E_{nl} -
    V_\mathrm{eff}(s)}\right) \;,
\end{equation*}
\cite{Ledoux2006} suggest to set it according to:
\begin{equation}
    \label{eq:r_max}
    \frac{1}{\epsilon}\int_{\max \mathcal{F}}^{R_\mathrm{max}}\text{d}s\sqrt{V_\mathrm{eff}(s) -
    E_\mathrm{max}} = 18 \;,
\end{equation}
such that $\partial_R u_{nl}(R) = 10^{-16} u_{nl}(R)$ and thus irrelevant under
double floating point precision. We adopt this strategy, i.e. find the outer turning
point $\max\mathcal F$ and then integrate outward until eq.~\eqref{eq:r_max} is satisfied.

\begin{figure}
    \centering
    \includegraphics[width=\columnwidth]{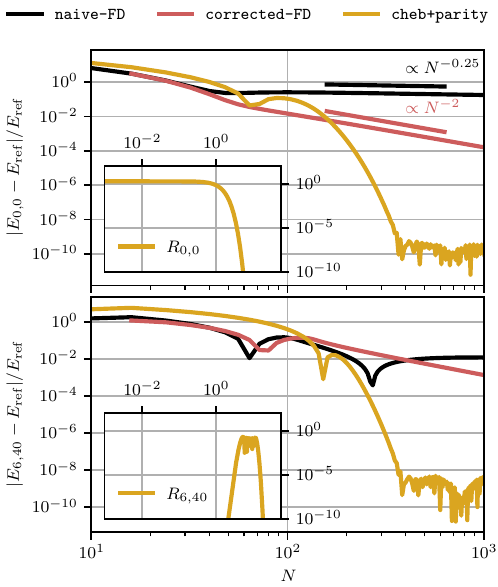}
    \caption{\label{fig:state_convergence}
        Convergence of the numerically computed eigenvalues $E_{nl}~=~E_{0,0}$
        \textbf{(top)} and
        $E_{nl}~=~E_{40,6}$ \textbf{(bottom)} for three different discretisation schemes.
        We find \texttt{naive-FD}, effective in the halo case, to show strongly
        degraded convergence properties under cylindrical symmetry. While the
        ad-hoc corrected finite difference scheme (see text) recovers quadratic
        convergence, it is outperformed by the exponential convergence of the
        parity aware Chebyshev pseudospectral approach \texttt{cheb-parity},
        which is the method of choice for this work. 
        \textbf{Insets:} High-resolution ($N=4096$) reference modes $R_{nl}$, 
        the eigenvalue of which is used as
        $E_\mathrm{ref}$.
    }
\end{figure}

It seems plausible to solve eq.~\eqref{eq:eigenproblem} via a
standard finite difference approximation, as is applicable in the halo case.
By approximating 
\begin{align*}
    \partial^2_R u_{nl}(R_k) &\approx D^2_R u_{nl}(R_k) \\
    &= \Delta R^{-2} \left[ u_{nl}(R_{k-1}) - 2
    u_{nl}(R_k) + u_{nl}(R_{k+1}) \right] \;,
\end{align*}
a tridiagonal, symmetric Hamiltonian is
recovered for which eigenvalues and vectors are easily computed.
Interestingly, this approach shows poor convergence properties at low angular momentum $l$ and most
pronounced for $l=0$. We depict the unsatisfactory situation in the upper panel
of Fig.~\ref{fig:state_convergence} in
terms of the convergence of the (specific) ground state energy as a function of the 
grid size $N$. Here \texttt{naive-FD} recovers an algebraic rate of convergence
of $\sim N^{-0.25}$ instead of the anticipated quadratic rate as is the
case under spherical symmetry.
Recall that the DF as shown in Fig.~\ref{fig:df} favours low angular momentum modes. Thus, accurate eigenfunctions in
this regime are critical.

Origin of the degraded convergence is the unsatisfactory approximation of the
second derivative for the small radius asymptotic behaviour of $u_{nl}$. As
$R \to 0$, eq.~\eqref{eq:eigenproblem} reduces to:
\begin{equation}
    \partial^2_R u_{nl} = \frac{l^2-1/4}{R^2}u_{nl}\,,
\end{equation}
and is solved by $u_{nl} \sim R^{l+1/2}$. While finite difference schemes
provide accurate approximations to power laws with integer exponents, rational exponents are
poorly recovered. In the present case, we have \citep{Laliena2018}:
\begin{equation}
\begin{split}
    D^2 u_{nl} 
    &= D^2 \left(\Delta R k\right)^{l + 1/2} 
    = W_l(k)\frac{u_{nl}}{R_k^2}\\
    &= k^2\left[\left(\frac{k-1}{k}\right)^{l+\frac{1}{2}} +
    \left(\frac{k+1}{k}\right)^{l+\frac{1}{2}} - 2\right]\frac{u_{nl}}{R_k^2}
    \neq \frac{l^2 - 1/4}{R_k^2} u_{nl}
    \;.
\end{split}
\end{equation}

Now, to make the physically correct asymptotic behaviour manifest at the
numerical level, it is permissible to simply substitute $l^2 - 1/4 \to W_l(k)$ in the
discretised equations yielding the improved scheme, \texttt{corrected-FD}, that
now recovers quadratic convergence to the ground state, cf. top panel of Fig
\ref{fig:state_convergence}.
\footnote{
    The astute reader may wonder, why we do not transform eq.~ \eqref{eq:eigenproblem}
    according to the discussion in Appendix~\ref{appendix:wkb} in order to
    recover a classical barrier shape $\propto l^2/R^2$ as this seems to
    recover a small radius power law asymptote with integer exponent.
    Unfortunately, if $l=0$ inhomogeneous Dirichlet conditions are
    imposed at the domain boundaries for a Langer corrected version of eq.~
    \eqref{eq:eigenproblem}. This generalises the discrete problem to an
    inhomogeneous eigenvalue problem that is not straight forward to solve.
}

Our numerical tests indicate that, while accurate at low angular momenta,
highly excited states with $l\geq40$ remain challenging for
\texttt{corrected-FD}, to the point that we recover spurious, degenerate eigenvalues 
--- impossible for bound states in a one dimensional potential. 
Our experiments suggest that the uniform grid paired with the low order derivative approximation hinders
quick convergence.

We therefore suggest the use of a Chebyshev pseudospectral approach, \texttt{cheb-parity}, 
that respects the parity properties of the untransformed radial modes $R_{nl}$ \citep{Lewis1990}.
Implementation details are beyond the scope of this work but can be found in \cite{Trefethen2000}.

Figure~\ref{fig:state_convergence} demonstrates the superior exponential
convergence rate of this method for both low $l=0$ and high $l=40$ states.
In practice, modes with angular momenta $l>100$ are robustly and efficiently
computable on moderate non-uniform grids of $N=1024$ points.

\section{WKB wave function coefficients}
\label{appendix:wkb}
The task is to establish an equivalence between the DF defined real space
density, cf. eq.~\eqref{eq:df_rho}:
\begin{align}
    \label{eq:df_rho_appendix}
    \rho_\text{DF}(R)
    &= \frac{4a^2}{R} \int\limits_{V(R)}^{\infty} \dd E  \hspace{-1.5em}
    \int\limits_{0}^{R\sqrt{2a^2(E-V_\mathrm{eff}(R))}}\hspace{-1.5em}\dd L 
                \frac{f(E, L)}{\sqrt{2a^2\left(E-V_\mathrm{eff}(R)\right)}}
                \;,\\
    \label{eq:Veff}
    V_\mathrm{eff}(R) &= V(R) + \frac{L^2}{2a^2R^2}
\end{align}
and the wave-counterpart, cf. eq.~\eqref{eq:self-consistency_filament}:
\begin{equation}
    \label{eq:rho_psi}
    \rho(R)  = \frac{\mu}{2\pi R} \sum_{n,l \ge  0}
    N_{l}|a_{nl}|^2|u_{nl}(R)|^2\;, \;
    N_l = \begin{cases}
        1, &l=0\\
        2, &l>0
    \end{cases} \;,
\end{equation}
confined to:
\begin{equation}
    \label{eq:Veff_psi}
    V_{\mathrm{eff},\psi}(R) = V(R)+\epsilon^2\frac{l^2-1/4}{R^{2}}\;,\quad
    \epsilon^2 = \frac{\hbar^2}{2m^2a^2}
    \;.
\end{equation}

For isotropic, spherical systems this has
been achieved in \cite{Yavetz2022} in the high-energy regime of the
WKB solution for $\psi$ --- an idea which we follow. We also use the opportunity to
emphasize a well-known, but sometimes overlooked,
detail of the WKB formalism for radial Schr\"odinger equations that is of
particular relevance at low angular momentum $l$. 

What's the WKB wave function $u_n$ of bound state $n$ in a regular, one-dimensional
potential $V(x)$? If $x \in \mathbb{R}$, i.e. defined on the entire real line, we have:
\begin{subequations}
    \begin{align}
    \begin{split}
    \label{eq:wkb_naive}
    u_{n}(x) &= \frac{2C}{\sqrt{p(x)}} \sin\left(\frac{1}{\hbar}\int_{x_1}^{x} \dd
    x' p(x') + \frac{\pi}{4}\right)
    \end{split} \;,\\
    \begin{split}
    \label{eq:wkb_wave_number}
    p(x) &=\hbar \sqrt{\frac{1}{\epsilon^2}\left(E_{n}-V(x)\right)}
    \end{split}\,,
    \end{align}
\end{subequations}
with normalisation constant $C$ and $x_1<x$ as turning point, where the effective momentum
satisfies $p(x_1) = 0$.

One can show \citep{Langer1937, Berry1973} that this result is incorrect in the presence of a domain
restricted to $R\in \mathbb{R}^+$ and an effective potential that diverges as
$R\to 0$ --- conditions both met for the radial Schr\"odinger equation
eq.~\eqref{eq:eigenproblem}, but also in the
spherically symmetric halo case
\footnote{
    For fully isotropic haloes \citep{Yavetz2022}, integrating out
    the angular momentum dependence renders the differences between eq.~
    \eqref{eq:wkb_naive} and the Langer-corrected wave function irrelevant.
}.
To be explicit, simply substituting eq.~\eqref{eq:Veff_psi} in eq.~\eqref{eq:wkb_wave_number} for $V(x)$, i.e.:
\begin{equation}
    p(R)=\hbar \sqrt{\frac{1}{\epsilon^2}\left(E_{nl}-V(R)-\frac{\epsilon^2(l^2
    - 1/4)}{R^2}\right)} \;,
\end{equation}
yields wave functions eq.~\eqref{eq:wkb_naive} which do not faithfully approximate the solutions to eq.~\eqref{eq:eigenproblem} \citep[see][for a selection of problems]{Berry1972}.

In the cylindrical setting relevant here, \cite{Berry1973} show that
transforming eq.~\eqref{eq:eigenproblem} via:
\begin{equation}
    x = \log R\;,\; u_{nl} = e^{x/2} R_{nl} \;,
\end{equation}
resolves all inconsistencies for states with $l>0$. The effect of above
transformation on the classical WKB result in eq.~\eqref{eq:wkb_naive} is
summarised by the simple substitution $l^2 - 1/4 \to l^2$ known as Langer's
correction \citep{Langer1937}. The corrected, effective momentum reads:
\begin{equation}
    \label{eq:Langer_momentum}
    p(R) = \hbar
    \sqrt{\frac{1}{\epsilon^2}\left(E_{nl}-V(R)-\frac{\epsilon^2l^2}{R^2}\right)}
    \;,
\end{equation}
which re-establishes the classical form of the angular momentum barrier in eq.~\eqref{eq:Veff} if we identify the trivial mapping:
\begin{equation}
    L_{nl} = l \times \hbar m^{-1} \;.
\end{equation}

\cite{Berry1973} show that s-waves ($l=0$) do not follow a Langer-corrected version of eq.~\eqref{eq:wkb_naive}.
However, at high energies, $E_{nl}\gg V$, correct WKB s-waves are asymptotically
equivalent to the $l>0$ modes. As this is our regime of interest, it is correct
to set $u_{nl}$
as eq.~\eqref{eq:wkb_naive}, but with the Langer corrected, effective momentum of
eq.~\eqref{eq:Langer_momentum} for all $l\geq 0$.

The normalisation constant $C$ follows from differentiation of the quantisation
condition  $\int_{R_1}^R\dd R' p(R') = (n+1/2)\pi\hbar$. One finds:
\begin{equation}
    C^2 = \frac{\hbar}{4\pi\epsilon^2} \frac{\dd E_{nl}}{\dd n} \;.
\end{equation}

Now, substituting eq.~\eqref{eq:Langer_momentum} into eq.~\eqref{eq:wkb_naive} and
subsequently into eq.~\eqref{eq:rho_psi}, converting the double sum into
integrals via $\dd L_{nl} = \hbar m^{-1} \dd l$ and $\dd E_{nl} = \frac{\dd
E_{nl}}{\dd n} \dd n$, and approximating $\sin^2(\cdot)\simeq 1/2$ (permissible due
to the fast oscillation of its argument) yields (under the assumption $E_{nl} \gg V$):
\begin{multline}
    \rho(R) \sim \frac{\mu}{2\pi^2\epsilon^2R}  \int \dd E_{nl} \dd L_{nl}
    \frac{|a_{nl}|^2}{\sqrt{2a^2\left(E_{nl}-V(R) -
    \frac{L_{nl}^2}{2a^2R^2}\right)}}\;.
\end{multline}
By comparing with eq.~\eqref{eq:df_rho_appendix}, we arrive at:
\begin{equation}
    |a_{nl}|^2 \sim \frac{(2\pi\hbar)^2}{m^2\mu} f\left(E_{nl}, L_{nl}\right)
    \quad (E_{nl} \gg V) \;.
\end{equation}

Above derivation is correct as long as $\rho_\mathrm{DF}$ and its associated DF
are bounded. This is of course violated for $\beta > 0$, cf. eq.~\eqref{eq:constant_beta_df}. No finite combination of eigenmodes will be able to
reproduce a cuspy density profile as all of them are bounded. 
The best we can do in this scenario is to reproduce a cuspy $\rho_\mathrm{DF}$ up to some 
length scale of interest. This is achieved by an ad-hoc change in the $l=0$ WKB
weights, if we set:
\begin{equation}
    L_{nl} = \max(L_0, l) \times \hbar m^{-1}\;,\; 0 < L_0 < 1\;.
\end{equation}

\section{Ellipsoidal collapse}
\label{appendix:ellipsoidal_collapse}
We briefly summarise some mathematical details of the ellipsoidal collapse
dynamics depicted in Fig.~\ref{fig:ellipsoidal_collapse}. An extensive
discussion of the model is given in \cite{Bond1996}. Our notation follows \cite{Angrick2010}.

We describe the evolution of the ellipsoid of mass $M$ in its principal axis system. Let
$a_i(a)$ be the scale factor in direction $i$, related to the physical size via
$r_i(a) = a_i(a) (\frac{3M}{4\pi\rho_m})^\frac{1}{3}$
and density contrast $\delta = a^3/(a_1a_2a_3) - 1$. Each axis then evolves under the influence of the
internal and external tidal forces and the cosmological background expansion.
Choosing a flat $\Lambda$CDM background with $E^2(a) = \Omega_\Lambda + \Omega_m
a^{-3}$ and density parameters $1=\Omega_m + \Omega_\Lambda$, one finds:
\begin{multline}
    \label{eq:ellipsoidal_collapse}
    0=\frac{\dd^2 a_i}{\dd a^2} + \left[\frac{1}{a} +
    \frac{E'(a)}{E(a)}\right]\frac{\dd a_i}{\dd a} +
    \left[\frac{3\Omega_m}{2a^5E^2(a)}C_i(a) -
    \frac{\Omega_\Lambda}{a^2E^2(a)}\right]a_i \;,\\
    \text{with} \;\;\;
    C_i(a) = \frac{1+\delta(a)}{3} + \frac{b_i(a)}{2} +
    \lambda_\mathrm{ext,i}(a) \;.
\end{multline}
Deviations from sphericity source tidal shears. Specifically,
\begin{equation}
    b_i(a) =
    a_1(a)a_2(a)a_3(a)\int_0^\infty\frac{\dd\tau}{\left[a_i^2(\tau)+1\right]\Pi_{k=1}^3\left[a_k^2(\tau)+1\right]^\frac{1}{2}}
    - \frac{2}{3} \;,
\end{equation}
represents the internal tidal shear, whereas:
\begin{equation}
    \lambda_\mathrm{ext,i}(a) = \frac{D(a)}{D(a_\mathrm{ini}}
    \left[\lambda_i(a_\mathrm{ini}) - \frac{\delta(a_\mathrm{ini})}{3}\right]
    \;,
\end{equation}
models the external tidal shear contribution.
Initial conditions for the scale factors $a_i$ at $a=a_\mathrm{ini}$ follow from
Zeldovich's approximation:
\begin{subequations}
    \begin{align}
        \begin{split}
            a_i(a_\mathrm{ini}) 
            = 
            a_\mathrm{ini}(1-\lambda_i(a_\mathrm{ini})) \;,
        \end{split}\\
        \begin{split}
            \frac{\dd a_i}{\dd a}\bigg\rvert_{a_\mathrm{ini}}
            = 
            1- \left(1-\frac{\dd \log D}{\dd\log a}\bigg\rvert_{a_\mathrm{ini}}\right)\lambda_i(a_\mathrm{ini})
        \end{split} \;.
    \end{align}
\end{subequations}
The eigenvalues of the Zeldovich deformation tensor $\lambda_i$ relate to the
shear ellipticity $e$ and prolaticity $p$. \cite{Sheth2001} show that
the most probable initial conditions have vanishing prolaticity, $p=0$, so that:
\begin{equation}
    \lambda_{1,3}(a_\mathrm{ini}) = \frac{\delta(a_\mathrm{ini})}{3}(1 \pm
    3e)\;,\quad
    \lambda_2(a_\mathrm{ini})= \frac{\delta(a_\mathrm{ini})}{3}\;.
\end{equation}
This concludes the specification of the initial conditions. What remains is a
freeze out condition for each principal axis. In accordance with the condition of
a quasi-virial filament state, cf. Sec.~\ref{subsub:real_space}, we adopt the freeze-out approach of
\cite{Angrick2010}, which applies the tensor viral theorem \citep{Binney2008} to
an ellipsoidal over density. One finds axis $i$ to be virialised once the criterion:
\begin{equation}
    \label{eq:freeze_out}
    \left(\frac{a_i'}{a_i}\right)^2 = \frac{1}{a^2 E^2(a)}\left(\frac{3\Omega}{2a^3}C_i -
    \Omega_\Lambda\right)\,,
\end{equation}
is met.

To compute the filament mass function, two ingredients are required: 
(i) the critical linear density, $\delta_\mathrm{ec}(e, z)$, at the redshift of filament formation,
and (ii) a mapping from ellipticity $e$ to filament mass $M$. 

For the first step, we determine the unique initial overdensity $\delta(a_\mathrm{ini})$ for a
given ellipticity $e$ that results in the second axis to collapse at redshift
$z$. To achieve this, we sweep over a broad range of ellipticities, optimize
$\delta(a_\mathrm{ini})$ to meet the filament formation condition, and then
extrapolate to $z$ via the linear growth factor $D(a)$. 

For the ellipticity-to-filament mass mapping, we proceed as follows: We first map
mass to the variance:
\begin{equation}
    \label{eq:variance}
    \sigma^2(R) = \int_0^\infty \frac{\dd k}{2\pi^2} k^2 P_\mathrm{FDM}(k) W(k\mid R)^2 \;,
\end{equation}
of the density field smoothed on spatial scales $R(M)$, and then map the
variance to the ellipsoid's ellipticity via the standard procedure outlined in \cite{Sheth2001}.
The linear matter power spectrum at $z=0$ is set according to the analytical transfer function 
of \cite{Hu2000}:
\begin{equation}
    P_\mathrm{FDM}(k) = T^2(k) P_\mathrm{CDM}(k)\;,\quad T(k) = 
    \frac{\cos\left((Ak)^3\right)}{1+(Ak)^8}\,,
\end{equation}
where $A=0.179 \left(m/\SI{1e-22}{\electronvolt}\right)^{-4/9}\mathrm{Mpc}$.
For the window function, $W$, a smooth-$k$ filter is used \citep{Du2023}:
\begin{equation}
    W(k \mid R) = \frac{1}{1+(kR)^\beta}\;,\quad 
    M(R)=\frac{4}{3}\pi(c_WR)^3\rho_m\;,
\end{equation}
with {\it N}-body calibrated values $\beta=9.10049$ and $c_W=2.1594$.

These steps provide us with the mass-dependent (or, equivalently, the variance-dependent via eq.~\eqref{eq:variance}) barrier shape:
\begin{equation}
    B(\sigma^2, z) = \frac{D(z)}{D(0)}\delta_\mathrm{ec}(\sigma^2,z)\;.
\end{equation}

The excursion set formalism postulates that a randomly walking overdensity in $\sigma^2$-space tracks the event of halo/filament formation (of mass $M(\sigma^2$)) as the moment when its trajectory passes $B(\sigma^2,z)$ for the first time. The multiplicity function, $f_B(\sigma^2)$, corresponds to the probability distribution of this event. Practically, it is found by carrying out Monte-Carlo simulations of the kind just described. For implementational convenience, however, we follow \cite{Zhang2006, Benson2013, Du2016} and express $f_B(\sigma^2)$ as the solution
to the integral equation:
\begin{equation}
    \int_0^{\sigma^2}\text{d}S\ f(S)
    \text{erfc}\left[\frac{B(\sigma^2,z) - B(S,z)}{\sqrt{2(\sigma^2 - S)}}\right]
    =\text{erfc}\left[\frac{B(\sigma^2,z)}{\sqrt{2\sigma^2}}\right]\;.
\end{equation}
Eq. \eqref{eq:fmf_def} then yields the FMF.

\section{Single cylinder power spectrum}
\label{appendix:cylinder_power}
An advantage of our bottom-up filament model is that we have access to a
semi-analytical version of the FDM wave function. Thus, we are able to compute
$P_\mathrm{single}(k)$ based off the eigenstate expansion directly rather than
populating a high-resolution box with $|\psi|^2$ and then treating it as a
generic density of which $k$-space histogram is computed.

To compute eq.~\eqref{eq:Psingle} we proceed as follows:
Firstly, Fourier transform the normalised density $u_\text{FDM}$, cf. eq.~\eqref{eq:u_FDM}:
\begin{equation}
\begin{split}
    \hat u_\text{FDM}(\bm k \mid M)
    &=
    \mathcal{F}_{3D}\left[u_\text{FDM}(\bm x \mid M)\right]\\
    &=
    \mathcal{F}_{3D}\left[M^{-1}\lambda\left(Z \mid L(M)\right)|\psi(R,
    \Phi)|^2\right] \\
    &= 
    M^{-1}\mathcal{F}_{1D}\left[\lambda\left(Z \mid L(M)\right)\right]
    \cdot \mathcal{F}_{2D}[|\psi(R, \Phi)|^2]
\end{split} \;.
\end{equation}
Our choice of longitudinal density, eq.~\eqref{eq:lambda}, was motivated by its rapidly
decaying, simple Fourier transform, 
\begin{equation}
    \mathcal{F}_{1D}\left[\lambda\left(Z \mid L(M)\right)\right] 
    =
    \frac{\sqrt{2/\pi} \exp\left(-\frac{k_Z^2}{4\Delta^2}\right)
    \sin\left(\frac{k_ZL(M)}{2}\right)}{k}\;.
\end{equation}
A simple closed form longitudinal spectrum reduces the problem of computing
$P_{\mathrm{single}}(k|M)$ to an effectively two-dimensional problem as 
$\hat u_\text{FDM}(\bm k \mid M)$ factorizes and its $k_Z$-component does not
need to be stored.

Secondly, to compute the cross-sectional contribution, $\mathcal{F}_{2D}[|\psi|^2]$, we note
that $\psi$ is periodic in $\Phi$. Thus, we may expand $|\psi|^2$ and it's
Fourier transform as:
\begin{equation}
    |\psi|^2(R,\Phi) = \sum_m f_m(R) e^{im\Phi}\;,\;\;
    \mathcal{F}_{2D}\left[|\psi|^2\right](k_R, \omega) = \sum_{m'} g_{m'}(k_R)
    e^{im'\omega}\;.
\end{equation}
Upon inserting the eigenstate ansatz for $\psi$ and using orthonormality of the
angular Fourier modes, we find:
\begin{equation}
    \label{eq:autocorrelation}
    f_m(R) = \braket{e^{im\omega} \mid |\psi|^2} = \sum_{n,n'}\sum_l
    a^*_{nl}R_{nl}(R)a_{n'l+m}R_{n'l+m}(R)\,,
\end{equation}
which is identical to the discrete autocorrelation of $a_{nl} R_{nl}(R)$ with
angular momentum lag $m$ and summed over all excitation numbers $n$.
The coefficient function $g_m(k_R)$ then follows from $f_m(R)$ via a
$m^{\text{th}}$-order Hankel transform $\mathbb{H}_m$ \citep{Baddour2009}:
\begin{equation}
    \label{eq:hankel}
    g_m(k_R) = 2\pi i^{-m} \int_0^\infty \text{d}R \; f_m(R)\,J_m(k_R R)\,R
    = 2\pi i^{-m} \mathbb{H}_m\left[f_m\right](k_R)\;
\end{equation}
--- conveniently implemented via the \texttt{FFTLog} algorithm \citep{Talman1978, Hamilton2000}
applied to each mode. This also fixes the values of $R$ to a log-uniform grid
on which eq.~\eqref{eq:autocorrelation} has to be evaluated.

Lastly, with $\hat u_\text{FDM}(\bm k \mid M)$ in \emph{cylindrical} coordinates
$\bm k = (k_R, k_Z, \omega)^\intercal$ at hand, we perform a \emph{spherical} average of the
squared modulus to arrive at the isotropic, single cylinder power spectrum:
\begin{equation}
\begin{split}
    P_\text{single}(k) 
    &= \left\langle |u(\bm k \mid M)|^2\right\rangle_{|\bm k| = k}\\
    &= \frac{1}{4\pi k^2} \int \text{d}^2\Omega(k)\;|u(\bm k \mid M)|^2\\
    &= \frac{1}{4\pi k} \int \text{d}\omega\,\text{d}k_Z\;\Big|u\left(\sqrt{k^2 - k_Z^2},
    k_Z, \omega \mid M\right)\Big|^2 \;.
\end{split}
\end{equation}
We repeat the outlined procedure for multiple wave function coefficient phase
realisations $i$ and then average all $P_{\text{single},i}(k)$ to arrive at an ensemble
estimate. The solid lines in both panels of Fig.~\ref{fig:powerspectra} show
this ensemble average.
\end{appendix}
\end{document}